\documentstyle{mn}
\title[Keck Spectroscopy of Faint Galaxies Identified as MicroJansky Radio Sources] {Keck Spectroscopy and Imaging of Faint Galaxies Identified as MicroJansky Radio Sources} 
\author[N.D. Roche, J.D. Lowenthal, D.C. Koo]{Nathan D.
Roche$^{1,4}$, James D. Lowenthal$^{2,5}$ and David C. Koo$^{3,6}$\\
$^1$ Institute for Astronomy,
 University of Edinburgh, Royal Observatory, Blackford Hill,
Edinburgh, EH9 3HJ, Scotland.\\
$^2$Department of Physics and Astronomy,
     University of Massachusetts,
     Amherst, 
     MA 01003-4525,
     USA.\\
$^3$UCO/Lick Observatory,
       Department of Astronomy and Astrophysics,
      University of California,
     Santa Cruz,
       CA 95064, 
       USA.\\
{$^4$ \verb"ndr@roe.ac.uk"}\hspace{8mm}   
{$^5$ \verb"james@velo.astro.umass.edu"}\hspace{8mm}
{$^6$ \verb"koo@ucolick.org"}\hspace{8mm}
}

\input{psfig.sty}

\begin{document}

\maketitle
\begin{abstract}
We investigate the nature of  the faintest radio sources detected in 3
VLA surveys, to  $\rm F(8.5~GHz)\sim 8\mu \rm Jy$.  Using the Keck Low
Resolution Imaging Spectrograph in $BRI$ and the Near Infra Red Camera
in  $K^{\prime}$ ($2.1\mu  \rm m$),  we  image 51  radio sources,  and
identify probable optical counterparts for 50.  With LRIS spectroscopy,
we  successfully acquire  new redshifts  for  17 sources.
Combining these with 9 prior redshifts, we can then analyse a sample 
of 26 $\rm \mu Jy$ sources with spectroscopic redshifts. 

Based on this sample of 26, we find the largest contribution, about 60
per cent (15), to be from  disk
 galaxies with high radio to optical ratios, indicating ongoing or
 very recent starbursts, at redshifts $z=0$--1.
 Most of these have large optical luminosites  
($M_B\simeq -22\pm
1$), although two are sub-$L^*$ ($M_B>-20$). About 20 per cent (5)  
of the sample have relatively low ratios of  radio
to optical flux, consistent with normal, non-interacting galaxies, all
at $z<0.4$. Four are QSOs, of which one 
appears to be interacting and  shows
strong MgII absorption. Two others
are very radioluminous giant ellipticals and presumably contain AGN.

All of the  14 non-QSO galaxies with spectra 
show  $H\beta$ and $\rm [OII]3727\rm \AA$ emission lines, with 
luminosities
  corresponding to star-formation rates (SFRs) of order 
$\sim 10 M_{\odot}$ $\rm yr^{-1}$. These SFRs are about an order
 of magnitude lower than estimated from the radio
 luminosities. This result may be explained if the galaxies are 
 observed $\sim 10^7$--$10^8$ yr after a starburst
and/or their regions of 
 current star-formation are heavily obscured by dust.
 Three of the galaxies show strong enough
 $\rm H\delta$
 absorption to be  classed as post-starburst or `e(a)' galaxies.

The Keck spectra are of high enough resolution to provide approximate
emission line widths and rotation curves, as  
well as  $\rm  [OIII]/H\beta$  ratios (`excitation').   Most  of  the
 galaxies (12/14) have high amplitude kinematics (100--$400$ km$\rm
 s^{-1}$), implying large dynamical masses,  and moderate excitation ($\sim 0.51\pm 0.12$),
 consistent with their morphological appearence as non-AGN, 
 $L\geq L^*$ spirals, of probable near-solar metallicity. 
 The two with low optical luminosities appear to be low-mass 
 galaxies with much higher excitations ($\sim 2.8$), and may be more similar
to local HII galaxies than normal spirals.

   \end{abstract}


\section{Introduction}
The  radio luminosity  of star-forming  galaxies is  a  combination of
thermal  emission  from  star-forming  regions,  discrete  synchrotron
emission  from  young  supernova  remnants,  and  diffuse  synchrotron
emission from  cosmic rays, which  also originate in  supernovae.  The
progenitors   of  supernovae   are   short-lived  $M\geq   8M_{\odot}$
stars. The  thermal component  of radio luminosity  will approximately
trace the instantaneous star-formation rate (SFR), and the synchrotron
component  the very  recent SFR,  within the  previous $\sim  10^8$ yr
(Condon  1992;  Lisenfeld  et  al.   1996). In contrast,
the  optical  and near  ultraviolet  fluxes  trace the  star-formation
integrated  over  longer  periods  and  are  very  sensitive  to  dust
extinction, which has no effect at radio wavelengths.

Deep radio surveys are  therefore useful in studying galaxy evolution,
and especially  the importance of  short-term, intense bursts  of star
formation.  Benn  et al. (1993)  found the majority of  faint, sub-mJy
sources  to be  star-forming  galaxies, rather  than  AGN.  Hammer  et
al.  (1995) investigated  the sources  detected in  a deep  Very Large
Array  (VLA) radio  survey  covering one  field  of the  Canada-France
Redshift  Survey to  a limit  $F(5.0~\rm GHz)\simeq  16\mu Jy$.  Of 36
sources, 34 had probable optical identifications and the CFRS provided
redshifts and spectra  for 23. The sources were found  to be a mixture
of  (i) ellipticals,  presumably  with AGN,  (ii)  disk galaxies  with
apparently post-starburst spectra, at redshifts out to $z=1.16$, and
(iii) lower redshift, high-excitation emission-line galaxies.
 
The  VLA  survey  of  Windhorst  et al.  (1995)  detected  sources  to
$F(8.5~\rm GHz)\simeq 8.8\rm \mu Jy$ on a field with HST WFPC imaging,
and optical  counterparts could be  identified for 21/22.  This sample
was found  to consist  primarily of disk galaxies. Of these, $\sim 60$
per cent  were in close  and probably interacting pairs,  implying
that  interaction-triggered starbursts  are the  dominant  mechanism in
producing the high radio luminosities. Locally, most
radio-selected starburst  galaxies are interacting  or merging (Smith,
Herter  and  Haynes  1998;  Serjeant et al. 2000),  as  are  almost  all
ultraluminous infra-red galaxies (ULIRGs) (Clements et al. 1996).
  
Richards  et  al. (1998)  detected  $F(8.5~\rm  GHz)\leq  6.3 \mu  Jy$
sources on  a VLA image  centered on the  Hubble Deep Field  (HDF), on
which redshifts have  been measured for most of  the brighter ($I<24$)
galaxies.   Again,  most   (70--$90\%$)  were   star-forming  spirals,
irregulars and  merging galaxies, over a wide $0<z<3$ redshift  range.
 The steep number counts and
extended $N(z)$ of galaxies
in deep radio surveys  indicate a strong evolution with redshift
in the total SFR (e.g. Haarsma et al. 2000).

In  this  paper,  we  investigate a sample of $\rm \mu Jy$  radio-detected
 galaxies,
 by performing a multi-passband imaging and spectroscopic survey of  
three deep VLA fields (including that of Windhorst et al. 1995).
We observe in the $B$, $R$ and (for some galaxies)
$I$  bands, and in the  near  infra-red
$K^{\prime}$ band. The high-quality  spectroscopy possible with LRIS
yields further information
on the nature of these galaxies.
 First, 
Lisenfeld et al. (1996) predict,  at the end of a starburst, the
thermal component of radio emission will decline within $\sim 20$ Myr
 but the
enhanced synchrotron emission will maintain a high radio luminosity
 for at least 
$\sim 70$ Myr. This means that if  starbursts are relatively brief
 ($<100$ Myr), a radio-selected sample would contain a significant
 fraction of  recent
post-starburst galaxies -- which could
be identified from their strong $\rm H\delta$ absorption
(e.g.  Delgado, Leitherer and Heckman 1999; Poggianti and Wu 2000). 

Second, the luminosities of some  emission lines (e.g. $\rm  H\beta$)  
trace the immediate SFR, and hence the ratio of line to 
 radio luminositity would be sensitive to dust extinction and/or
 starburst
 age.  
Third, the ratio of the $\rm [OIII]5007\AA$ and $\rm H\beta$ 
fluxes (`excitation') is sensitive to the Hubble type and metallicity of a 
galaxy and may reveal an AGN contribution (Tresse et
al.  1996; Sodre and Stasinska 1999). 
 Fourth, the high spectral resolution will enable us to
 derive kinematic information from the 2D (spatial/spectral)
 emission-line profiles.

Section 2 gives details of our data, and Section
3  their reduction.  Section  4 describes  the  identification of  the
optical  counterparts  of  the  radio  sources,  lists  positions  and
magnitudes  and shows optical  images. Section  5 presents  our source
spectra, and  Section  6  discusses  optical and  radio  luminosities  and
colours.  In Section 7  we identify the  galaxies in  close  or interacting
pairs, in Section 8 interpret the spectral line equivalent widths, and in
Section  9  use line  profiles  to investigate galaxy
kinematics.  Section 10  is  an  overview and  discussion  of all  our
results on $\sim 10\mu \rm Jy$ sources.

Throughout, magnitudes are  given in the AB system,  defined such that
$m_{AB}=48.60-2.5~{\rm  log}_{10}~f_{\nu}$ where  $f_{\nu}$ is  in erg
$\rm    s^{-1}Hz^{-1}$,    or,    equivalently,    $f_{\nu}=3631\times
10^{-0.4m_{AB}}$    Jy.   For    the    passbands   of    observation,
$B_{AB}=B_{Vega}-0.048$,                       $R_{AB}=R_{Vega}+0.175$,
$I_{AB}=I_{Vega}+0.483$,
$K^{\prime}_{AB}=K^{\prime}_{Vega}+1.791$. Quantities dependent on
Hubble's Constant are given in terms of $h_{50}=H_0/50$ km $\rm s^{-1}$. 

\section{Observational Data}
\subsection{Radio Surveys}
This study is  based on the published source  catalogs from three deep
radio  surveys, of fields  originally studied  in the  Leiden Berkeley
Deep  Survey  (Windhorst, van  Heerde  and  Katgert  1984). The  radio
observations  were  carried out  using  the  National Radio  Astronomy
Observatory Very  large Array (VLA)  in San Augustin, New  Mexico. The
first  survey is of  the `Lynx2'  field, centered  at R.A.  $08^h 45^m
04^s$  Dec   44:34:05  (equinox   2000.0  is  used   throughout).  VLA
observations of  63 hours in December  1989 and January  1990, at 8.44
GHz, gave  a noise level of  $\sigma=3.21\rm \mu Jy$  and an estimated
$14.5\rm \mu Jy$ completeness limit. Windhorst et al. (1993) cataloged
46 sources detected at $\geq 4\sigma$ , and we adopt their source
numbers (prefixed `16V'). Of these, only 14 belong to a `complete
sample' which have a peak flux density $\geq 14.5\rm \mu Jy$
and are  within 4.59 arcmin of the field centre. The
names of sources not in the complete sample are suffixed with `*'. The
source `16V29A*' was considered to be part of the same extended object
as 16V29, and here the fluxes are combined into a single 16V29 source.

The second, and  deepest, VLA survey is centered  on the `Lilly field'
of  the   HST  Medium   Deep  Survey,  R.A.   $31^h  12^m   17^s$  Dec
+42:38:06.  VLA observations, again  at 8.44  GHz, totalled  159 hours
between  October   1993  and  January  1995,  giving   a  noise  level
$\sigma=1.5\rm \mu Jy$ and an estimated completeness limit $8.8\rm \mu
Jy$.  Windhorst et el.  (1995) list 24  sources and identify
their  optical counterparts  on  WFPC images;  we  adopt their  source
numbers, adding a prefix `L'. However, two of these (L14 and L22) are
actually  non-detections and not  considered here.  We retain  L20 and
L21, which  are detections  but have fluxes  7--$8.8\rm \mu  Jy$, just
below the completeness limit.

Thirdly, we  include some sources from  the larger area  but less deep
survey  of Weistrop  et al.  (1987), in  which a  field known  as SA68
(R.A. $23^h  59^m 15^s$ Dec 14:55:00)  was observed at 5.0  GHz for 40
hours,  detecting 9  sources to  a completeness  limit $60\rm  \mu Jy$
($\sim 40$--$45\rm  \mu Jy$ at 8.44  GHz). We adopt their nomenclature,
abbreviating the `SA68' prefix to `S'. 

The VLA  surveys of  Lynx2 and  SA68 have only  FWHM $\sim  10$ arcsec
resolution -- hence, although Windhorst et al. (1993) considered
most  Lynx2 source  positions accurate  within $<2$  arcsec,  we might
expect a few to have offsets  of up to $\sim 7 $ arcsec. On the
Lilly field the use of  VLA C-configuration improved the FWHM to $\sim
3$ arcsec.
\subsection{Optical Imaging}
Our optical imaging observations were performed using the Low
Resolution Imaging Spectrometer  (LRIS), on the  Keck telescope  in
Hawaii (Oke et al. 1995).  This camera,
is fitted with a $2048\times 2048$ pixel CCD, giving a $\sim 7\times 7$ arcmin
field-of-view with  a pixelsize  0.215 arcsec. LRIS  observations were
taken  on the  3rd  and  4th of  January  1995, and  made  use of  $B$
($\lambda_{eff}=4377\rm     \AA$,    FWHM    $919\rm     \AA$),    $R$
($\lambda_{eff}=6417\rm   \AA$,   FWHM    $1185\rm   \AA$)   and   $I$
($\lambda_{eff}=8331\rm \AA$, FWHM $3131\rm \AA$) filters.

Pointing  at R.A. $08^h 45^m 00^s$ Dec 44:34:59, we imaged
the central part of the Lynx2 field, for
 $8\times 300$  sec exposures in $B$, $11\times  300$ sec in
  $R$ and $6\times  300$ sec in $I$.  The Lilly  field, where the LRIS
  CCD covers the whole area of the Windhorst et al. (1995) survey, was
  observed for  $2\times 300$ sec  in each of  $B$ and $R$.   The SA68
  radio sources are more widely spread and it was necessary to observe
  them  individually.  A field  centered  on  SA68:10  was imaged  for
  $2\times 300$ sec in $B$, $4\times 300$ sec in $R$ and $3\times 300$
  sec in $I$, and three other fields in $I$ only, for a single 300sec
  exposure -- the first containing SA68:02 and SA68:04, the
second SA68:09 and the third SA68:12.

\subsection{Near Infra-Red Imaging}
Using the Keck telescope  Near Infra-Red Camera (NIRC) (Matthews and
    Soifer 1994), we observed the sources in the 
    the $K^{\prime}$ passband ($\lambda_{eff}=2.124\rm \mu m$ FWHM
  $0.337\rm \mu m$), on
		  the 8th of      January 1996. NIRC
   is fitted with a $256\times 256$ pixel indium-antimonide (InSb)
 detector, providing a 37.5 arcsec field of view with pixelsize 0.15
  arcsec. Most sources were observed for 5 consecutive
spatially-dithered exposures of 60 sec (each exposure was in turn made
 up of 6 co-added 10 sec integrations); a few received 10 exposures.
  \subsection{Optical Spectroscopy} Spectroscopic observations were
performed using LRIS, on the same nights as the
			  direct imaging, and made use of
    multi-slit masks or, for a few galaxies, single long
slits. The spectrograph is sensitive at 4000--$10000\rm \AA$, but only
 a $\sim 2600\rm \AA$ range is observed for each source, the starting
 wavelength depending on the slit position. With the  600 lines $\rm
  mm^{-1}$  grating, the wavelength dispersion is $\rm 1.27\AA$ $\rm
pixel^{-1}$ with resolution $3.1\rm\AA$ FWHM.

 The  Lynx 2  field multi-slit  mask had  22 slits,  of which  12 were
positioned on  likely optical counterparts  for the radio  sources and
the remainder  on other galaxies  for a separate redshift  survey, and
received  $11\times  1000$  sec  exposures. Four  Lynx2  sources  were
observed  with a  single long  slit  and either  one or  two 1800  sec
exposures.  The  Lilly  field   multi-slit  mask  had  11  slits,  all
positioned  on   radio  sources,  and  received   $3\times  1800$  sec
exposures. The long slit was again  used for five of the SA68 sources,
with 1000 or 1800 sec exposures.
\section {Data Reduction and Analysis}
\subsection{Optical Data}
The LRIS images were debiased, and flat-fielded using `pyraf' routines
   developed by Dan Kelson. Multiple exposures were averaged, with
  rejection of cosmic rays using the `nukecr' routine of Luc Simard.
 The stacked $B$, $R$ and $I$ frames of the Lynx2 field, and the $B$
and $R$ frames of the Lilly field, were then positionally registered.

 Sources  were detected  and catalogued  using SExtractor  (Bertin and
 Arnauts 1996) in `double-image mode',  meaning that one frame is used
 for  source detection  but  the  fluxes are  measured  from the  same
 positions  on  a second.   For  each of  the  fields  we coadded  the
 exposures in  different passbands (weighting  each by the  inverse of
 its rms) to give combined `BRI'  and `BR' images, which were used for
 source  detection. SExtractor  measured 'Kron-type'  total magnitudes
 from  the single-passband images  within elliptical  apertures, which
 are fitted  to each source on  the detection frame.  The  aim of this
 was  (i)  to  optimize  the detection  of  faint  sources  by
 combining all the data, and  (ii) to obtain more reliable colours, by
 positionally  matching the  photometric apertures  used  in different
 passbands.

For the  Lynx2 field, the  usable area covered  in all 3  passbands was
$5.03\times  6.93$  arcmin.  The  resolution was  estimated,  from  the
SExtractor FWHM of objects classed  as stars, as $1.34\pm 0.05$ ($B$),
$1.45\pm  0.03$ ($R$)  and $1.23\pm  0.03$ ($I$)  arcsec, and  the sky
noise  $\sigma_{sky}$ as $26.97(B)$,  $27.23(R)$, $26.63(I)$  mag $\rm
arcsec^{-2}$.  Our Lilly field image covered $5.63\times 7.21$ arcmin,
with somewhat poorer resolution, estimated as $3.03\pm 0.11$ ($B$) and
$2.07\pm  0.11$  ($R$)  arcsec,  and  is  a  little  less  deep,  with
$\sigma_{sky}$ of $26.39(B)$ and $26.47(R)$ mag $\rm arcsec^{-2}$.

Sources were detected as areas of 14 or more contiguous pixels above a
threshold  of  $1.0\sigma_{sky}$  (Lynx2 field)  or  $1.1\sigma_{sky}$
(Lilly  field)  above  the  sky  background on  the  detection  frame.
Detections  were  deblended  with  a  minimum  contrast  threshold  of
0.02. This gave  a total of 2462  detections on the  Lynx2 field, with
source counts turning over at $B>26.5$, $R>26$ and $I>26$, and 1506 on
the Lilly  field, with counts  turning over at $B>25.5$  and $R>25.5$.
For the  SA68:10 field, the  $B$ $R$ and  $I$ frames were  coadded and
sources  detected   above  a  $1.25\sigma_{sky}$   threshold.  The  $I$-band
exposures  of the  other three  SA68 sources  were analysed  as single
images, again with $1.25\sigma_{sky}$ thresholds.

Star-galaxy classification was performed using plots of FWHM against
magnitude, on which the stellar locus was separable from the
larger  galaxies  to $I=20.4$  on  Lynx2  and  $B=21.2$ on  the  Lilly
field. Fainter detections were classed as galaxies.

\subsection{Near Infra-Red Data}
The  NIRC $K^{\prime}$  images  were reduced  using {\sevensize  IRAF}
`dimsum' (created by P.  Eisenhardt, M. Dickinson et al.), which
positionally registered and  stacked the dithered exposures, rebinning
to  0.0375 arcsec  pixels, while  subtracting the  sky  background and
removing  cosmic  rays.  The  resolution  of the  stacked  frames  was
estimated  from  stellar images  as  0.74  arcsec FWHM,  significantly
better than the $BRI$ data. Objects were detected and catalogued using
SExtractor   with  a   minimum  area   16  pixels   and   a  threshold
$2.25\sigma_{sky}$    ($\sigma_{sky}=21.7$--22.4    mag   $\rm
arcsec^{-2}$), which typically gave 5--10 detections per NIRC frame.

\subsection{Photometric Calibration}
Photometric zero points were calculated for the Lynx2 $B$, $R$ and $I$
images using $\sim 50$  previously photometrically catalogued stars on
the observed field.   These zeropoints were then applied  to the Lilly
and  SA68 data  with corrections  for the  differences  in observation
airmass  (assuming  $\Delta   B=0.233X$,  $\Delta  R=0.067X$,  $\Delta
I=0.014X$  from a  tabulated extinction  curve). Magnitudes  were also
corrected  for   Galactic  extinction,  as  given   by  the  NASA/IPAC
Extragalactic Database (in the  $B$-band, 0.112 mag for Lynx2, 0.058
for the  Lilly field and  0.226 for SA68).  The  $K^{\prime}$ zeropoint
was determined from two standard stars, FS 5 ($K^{\prime}=13.342$) and
FS  21  ($K^{\prime}=13.115$), observed  on  the  same  nights as  the
galaxies.

Photometric errors in $R$,  $I$
and $K^{\prime}$ magnitudes are generally $\leq 0.1$ mag for the brighter,
 $R< 23.5$, galaxies. In the $B$-band the errors are
$\leq 0.1$ mag to $B\sim 22.8$ but for  
fainter $B$ magnitudes may exceed  $\sim 0.2$ mag.

\subsection{Spectroscopic Data}
Reduction of  the LRIS  multi-slit and single-slit  spectroscopic data
began with  debiasing and  cleaning of bad  pixels, using  the `pyraf'
routines   of  Dan  Kelson.   Multiple  exposures   were  positionally
registered  to the  nearest  integral pixel,  and cosmic-rays  removed
using `nukecr'.  The `rmydist' routine of  Dan Kelson was  used to fit
and  remove   (by  rebinning)  the  slight   $y$  (spatial)  direction
distortion of LRIS spectra.

The first exposure of the Lynx2  mask was divided into 22 strips, each
containing the light  from one of the spectrograph  slits. Each of the
strips was  analysed separately, following essentially  the methods of
Mallen-Ornel\'{a}s (2000). A wavelength  calibration was fitted to the
sky emission lines  on each strip, with rms  residuals of 0.5--$0.7\rm
\AA$.   LRIS spectra  also suffer  a $y$-dependent  distortion  in the
$x$-direction,  resulting in  the  sky lines  being  slanted from  the
vertical. To correct for this, the wavelength calibration was refitted
to  the  sky  lines  at   several  $y$  positions  on  each  strip,  a
two-dimensional function was then refitted  
with {\sevensize IRAF} `fitcoords',
and ({\sevensize IRAF} `transform') was used 
to rebin  each line of  the spectra to  a common and  linear $\lambda$
scale.

This transform  was then  applied to the  subsequent exposures  of the
same spectrum. There  were sometimes small ($\sim  1\rm \AA$) offsets in
$\lambda$  between  the transformed  exposures.  These were  measured
using {\sevensize IRAF} `xregister', and the pre-transform strips were
then re-transformed,  with the offsets  subtracted from the  $x_1$ and
$x_2$ parameters. This reduced offsets between exposures 
to  $<0.2\rm \AA$.   The same  procedure was  then applied  to  the 11
strips  of the multi-slit  Lilly field  mask and  to the  much broader
strips  of  the  long-slit  images.  The multiple  exposures  of  each
spectrum were  combined by  simple averaging, and  then sky-subtracted
using  {\sevensize  IRAF} `background'  --  at  each  position on  the
wavelength  ($x$)  axis,  4th-order   functions  were  fitted  to  the
variation of  the sky  in the spatial  ($y$) direction  (excluding the
region where the galaxy spectrum  is visible), and then subtracted, to
give a sky-subtracted 2D  spectrum.  One-dimensional spectra were then
extracted   by   using   {\sevensize   IRAF}{   `apall'   to   perform
(variance-weighted)  extractions in apertures  of width  12--20 pixels
(2.6--4.3 arcsec),  centered on each galaxy spectrum  and covering its
full width in the spatial  direction.  The extracted spectra are shown
in Section 5 below.

\section{Optical Counterparts of the Radio Sources}
\subsection{Identification}
The areas covered by our LRIS and/or NIRC images contain a total of 51
catalogued VLA  sources, of which 40  are in a  `complete sample' (the
16V sources listed as  being in the complete sample, plus  all the
others  except L20 and  L21).  Many of  these coincide
obviously  with a  bright  galaxy, but  in  other cases  there may  be
several fainter  galaxies within the `error box'.   We determine which
galaxies are the most probable identifications as described below.

(i) For  both Lynx2 and Lilly  fields, we used the  positions of faint
stars from  the USNO catalog  -- of which  $\sim 20$ were  detected on
each  frame area  - with  the  {\sevensize IRAF}  `pltsol' routine  to
derive accurate astrometric transforms,  with RMS residuals $\sim 0.6$
arcsec. 

 (ii) These astrometric transforms were used to convert the centroid positions
of the  radio sources  (Windhorst et al  1993, 1995) to
pixel co-ordinates on the optical CCDs. 

  (iii) Using  SExtractor, an
integral  number  count  $n(m)$  of  objects classed  as  galaxies  is
determined for  each LRIS  field, as a  function of  optical magnitude
(for all passbands used). 

 (iv)  For each  galaxy detected  within 7
arcsec of one of the radio sources, a probability,
$$P=1.0-{\rm  exp}(-\pi r^2  n(m))$$ is  estimated, where  $r$  is the
distance from  the radio source  centroid to the galaxy  position, and
$n(m)$  the  integral  number  count  (surface  density)  of  galaxies
brighter than  the galaxy under consideration (magnitude  $m$). $P$ is
then the  probability that a  galaxy of similar or  greater brightness
would, entirely  by chance, lie at  a similar or  closer distance from
the radio  source. The `best'  optical counterpart is then  the galaxy
with the lowest $P$.

 The  analysis  was  repeated   with  $n(m)$  from  different  optical
 passbands. Galaxies with  $P<0.05$ in one or more  passbands are then
 considered  to be  very probable  optical  IDs. For  each source  the
 reliability  of  the identification  is  represented  as  ID class  I
 ($P<0.05$),  II ($0.05<P<0.1$)  or III  ($P>0.1$), where  $P$  is the
 lowest value of the 2 or 3 passbands.

In total  we find  39 IDs of  class I  (with $\langle r  \rangle= 1.4$
arcsec), 7 of class II ($\langle r \rangle= 3.9$ arcsec) and 4
of class III ($\langle r  \rangle= 4.7$ arcsec). Using our astrometric
transforms,  the   optical  ID  positions  are   converted  to  RA/Dec
co-ordinates,  listed in  Table  1 together  with  the radio  centroid
co-ordinates from the VLA data.  Only  one of the 51 sources, L12, had
no detectable  counterpart within  7 arcsec of  the radio  position on
either the  LRIS or NIRC images.  Windhorst et al.  (1995) thought L12
might be  part of the  large L13 galaxy,  and we do not  consider this
`source' any further.

To  investigate the  reliabilty  of our ID technique,  we
performed  simulations in  which  radio sources  were assigned  random
positions   against    a   background   consisting    of   the   Lynx2
galaxies. Simulated optical counterparts  were then added with optical
magnitudes and separations from  the radio sources randomly drawn from
the magnitudes and separations of the `class I' IDs. 
The simulated radio  sources were then  `$P$-tested' against
catalogs consisting of the background galaxies plus the
 simulated counterparts.
\onecolumn
\begin{table}
\caption{Co-ordinates (equinox 2000.0) of the VLA radio sources and of
the  galaxies  identified   as  most  probable  optical  counterparts,
separation $r$  of the radio  and optical positions,  classes (I--III)
for the reliability of the ID, 8.44 GHz fluxes, magnitudes (AB system)
and approximate morphological types  of the galaxies (d=disk, b=bulge,
i=irregular, Q=QSO, {\it p}=pair {\it d}=disturbed.}
\begin{tabular}{lccccccccccc}
\hline   Radio  &   \multispan{2}   \hfil  Radio   position  \hfil   &
\multispan{2}\hfil Optical position \hfil & $r$~~ID & 8.44 GHz & $B$ &
$R$ & $I$ & $K^{\prime}$ & Mo. \\
\smallskip
Source & R.A. & Dec. & R.A.  & Dec. & arcsec~class & flux $\rm \mu Jy$
& mag & mag  & mag & mag & type \\ 16V08  & 8:44:43.90 & 44:34:01.22 &
8:44:43.75 & 44:33:59.74 & 2.2 ~ II & $58.0\pm 14.3$ & 24.45 & 24.53 &
24.57 &  - &  d \\ 16V10*  & 8:44:45.90  & 44:33:03.21 &  8:44:45.56 &
44:32:59.82 &  5.0 ~ III &  $76.1\pm 23.1$ &  25.28 & 24.27 &  23.18 &
20.48  &  d  \\  16V13  &  8:44:50.39 &  44:35:55.37  &  8:44:50.62  &
44:35:54.36 & 2.6 ~ I & $62.1\pm 15.8$ & 22.54 & 22.16 & 21.66 & 20.51
&  Q{\it  p} \\  16V15*  & 8:44:53.30  &  44:32:37.90  & 8:44:53.24  &
44:32:39.11 & 1.4 ~ I & $21.4\pm  5.6$ & 23.01 & 21.90 & 21.56 & 19.62
&  d{\it  p} \\  16V16*  & 8:44:56.68  &  44:33:17.01  & 8:44:56.61  &
44:33:15.68 & 1.6 ~ I & $14.5\pm  3.9$ & 25.27 & 23.86 & 22.88 & 20.25
& d \\  16V17 & 8:44:56.82 & 44:34:51.11 &  8:44:56.82 & 44:34:51.26 &
0.2 ~ I & $20.2\pm 4.3$ & 18.42  & 18.58 & 19.21 & 18.52 & Q \\ 16V18*
& 8:44:56.91  & 44:37:44.70  & 8:44:56.73  & 44:37:43.76 &  2.3 ~  I &
$75.3\pm 23.8$  & 22.94 &  23.69 & 23.80  & - &  d{\it p} \\  16V19* &
8:44:56.80  & 44:31:35.11  &  8:44:56.98 &  44:31:32.44  & 3.3  ~ I  &
$24.8\pm 6.8$ & 22.10 & 22.35 & 22.02 & - & d \\ 16V20* & 8:44:57.19 &
44:34:31.69 &  8:44:57.32 & 44:34:31.62  & 1.3 ~  I & $14.2\pm  3.8$ &
24.70 & 22.30 & 21.20 & 19.25  & d \\ 16V21 & 8:44:57.82 & 44:33:15.75
& 8:44:58.22 & 44:33:13.46 & 4.9 ~ I & $50.4\pm 5.5$ & 22.59 & 22.98 &
23.09  & 23.07  &  d{\it p}  \\  16V22 &  8:44:59.03  & 44:33:49.28  &
8:44:59.00 & 44:33:44.00 & 5.3 ~ II  & $20.6\pm 3.6$ & 23.18 & 22.65 &
22.59 & 22.53 & d \\  16V23* & 8:44:59.30 & 44:35:29.66 & 8:44:59.36 &
44:35:30.15 & 0.8 ~ I & $19.2\pm 4.6$  & 23.51 & 22.67 & 21.64 & - & d
\\ 16V24 & 8:44:59.93 & 44:33:49.83 & 8:45:00.28 & 44:33:49.20 & 3.7 ~
II &  $32.2\pm 3.7$ &  23.72 & 24.15  & 23.85 & 23.91  & d \\  16V25 &
8:45:05.48  & 44:34:15.62  &  8:45:05.48 &  44:34:15.14  & 0.5  ~ I  &
$31.5\pm  3.6$ &  24.91  &  22.92 &  21.89  & 19.46  &  d  \\ 16V26  &
8:45:05.57  & 44:33:58.62  & 8:45:05.61  & 44:33:56.56  & 2.1  ~  II &
$35.3\pm  3.6 $  & 24.91  & 25.39  & 24.75  & 21.54  & d  \\  
16V28 &  8:45:06.17 &  44:35:19.68 &  8:45:06.13 &
44:35:19.38 & 0.6 ~ I & $110.0\pm 10.0 $ & 15.94 & 15.36 & 15.68 & - &
$\rm d{\it p}^{[1]}$ \\ 16V29  & 8:45:06.32 & 44:36:24.68 & 8:45:06.04
& 44:36:23.93 & 3.1 ~ I & $155.6\pm 24.3 $ & 16.43 & 15.75 & 15.85 & -
&  d{\it  p} \\  16V30*  & 8:45:07.77  &  44:33:49.69  & 8:45:07.91  &
44:33:51.60 & 2.4 ~ I & $13.8\pm 3.4 $ & 22.29 & 20.94 & 20.33 & 18.37
&  d{\it  p}  \\ 16V31  &  8:45:08.30  &  44:34:40.37 &  8:45:08.35  &
44:34:38.58 & 1.9 ~ I & $62.4\pm 5.1 $ & 22.86 & 21.44 & 20.90 & 18.73
& d \\  16V34 & 8:45:14.46 & 44:34:51.53 &  8:45:14.33 & 44:34:50.97 &
1.6 ~ I & $31.0\pm 5.9 $ & 21.28 & 20.29 & 19.90 & 18.10 & d{\it d} \\
16V35* & 8:45:17.07  & 44:35:18.58 & 8:45:16.92 &  44:35:24.52 & 6.2 ~
II &  $25.4\pm 7.5 $ & 22.84  & 23.07 & 22.86  & 22.49 & d  \\ 16V36 &
8:45:19.28  & 44:46:46.26  &  8:45:19.49 &  44:35:42.87  & 4.1  ~ I  &
$78.10\pm 14.90$  & 20.71 & 19.39  & 19.00 &  17.40 & d{\it p}\\  L1 &
13:12:14.45  & 42:38:21.6  & 13:12:14.55  & 42:38:21.52  & 1.1  ~  I &
$9.8\pm 2.0  $ & 23.08  & 23.31 &  $22.23^{[2]}$ & 20.59  & d \\  L2 &
13:12:14.51  & 42:37:31.1  & 13:12:14.63  & 42:37:31.47  & 1.4  ~  I &
$8.8\pm  2.0 $  & 25.27  &  25.15 &  $25.62^{[2]}$ &  -  & d  \\ L3  &
13:12:15.08  & 42:37:02.6  & 13:12:15.15  & 42:37:01.84  & 1.1  ~  I &
$12.1\pm 2.0 $ &  20.71 & 20.19 & $21.13^{[2]}$ & 19.39  & d{\it p} \\
L4 & 13:12:15.29 & 42:39:00.9 &  13:12:15.23 & 42:39:00.56 & 0.7 ~ I &
$27.3\pm  2.0 $  & 17.87  & 18.45  & $18.21^{[2]}$  & -  & Q  \\  L5 &
13:12:16.07  & 42:39:21.3 &  13:12:16.49 &  42:39:19.9 &  4.9 ~  III &
$31.8\pm  2.0 $  & 25.09  & 24.18  & $24.82^{[2]}$  & -  & d  \\  L6 &
13:12:17.17  & 42:39:12.5  & 13:12:17.19  & 42:39:11.72  & 0.8  ~  I &
$23.2\pm 2.0 $ &  21.94 & 20.85 & $20.47^{[2]}$ & 18.78  & d{\it p} \\
L7 & 13:12:17.59 & 42:39:30.6 &  13:12:17.36 & 42:39:26.99 & 4.5 ~ III
& $15.2\pm 2.0$  & 25.58 & 25.57 &  $25.22^{[2]}$ & 21.09 & d  \\ L8 &
13:12:18.30  & 42:39:08.1  & 13:12:18.28  & 42:39:08.82  & 0.8  ~  I &
$15.3\pm 2.0$ & 25.88 & 25.24 & $25.62^{[2]}$ & 23.13 & d{\it p} \\ L9
& 13:12:18.46  & 42:38:43.9 &  13:12:18.44 & 42:38:43.50  & 0.5 ~  I &
$21.8\pm 2.0$  & 23.20 & 21.86 &  $20.70^{[2]}$ & 18.68 &  d{\it p} \\
L10 & 13:12:19.57 & 42:38:33.0 & 13:12:19.58 & 42:38:32.80 & 0.3 ~ I &
$17.8\pm 2.0 $  & 22.27 & 21.42 &  $20.87^{[2]}$ & 19.60 & d  \\ L11 &
13:12:20.16  & 42:37:03.4  & 13:12:20.23  & 42:37:04.33  & 1.2  ~  I &
$9.6\pm 2.0$ & $28.81^{[3]}$  & $25.03^{[3]}$ & $>25.72^{[2]}$ & 22.37
& d \\ L12 & 13:12:20.02 & 42:39:23.4  & [4] & & - ~ - & $11.8\pm 2.0$
& [4] & [4] & [4] & [4] &  - \\
 L13 & 13:12:21.04 & 42:39:23.3 & 13:12:21.02 &
42:39:22.26 & 0.3 ~ I &  $10.1\pm 2.0$ & 19.69 & 18.80 & $18.40^{[2]}$
&  17.19  & d  \\  L15  & 13:12:21.33  &  42:37:22.9  & 13:12:21.31  &
42:37:22.69 & 0.3 ~ I &  $10.1\pm 2.0$ & 19.34 & 18.68 & $18.22^{[2]}$
&  17.54  & d  \\  L16  & 13:12:21.83  &  42:38:27.3  & 13:12:21.82  &
42:38:26.68 & 0.6 ~ I &  $16.6\pm 2.0$ & 21.80 & 21.88 & $23.14^{[2]}$
& 21.52 &  d{\it p} \\ L17 & 13:12:22.49 &  42:38:14.1 & 13:12:22.39 &
42:38:13.62 & 1.3 ~ I & $11.5 \pm 2.0$ & 20.65 & 20.44 & $19.99^{[2]}$
&  19.56  & Q  \\  L18  & 13:12:23.24  &  42:39:08.6  & 13:12:23.22  &
42:39:07.42 & 1.2 ~ I & $22.6 \pm 2.0$ & 21.82 & 20.87 & $19.71^{[2]}$
& 18.04 &  d{\it p} \\ L19 & 13:12:23.67 &  42:37:11.9 & 13:12:23.70 &
42:37:11.60 & 0.4 ~ I & $24.4 \pm 2.0$ & 20.94 & 19.70 & $19.18^{[2]}$
& 17.68 &  d{\it p} \\ L20 & 13:12:24.11 &  42:38:05.2 & 13:12:23.74 &
42:38:06.69  &  4.5  ~  III  &  $8.3\pm  2.0  $  &  24.50  &  24.53  &
$24.82^{[2]}$ & 21.10  & d{\it p} \\ L21 &  13:12:24.42 & 42:37:05.2 &
13:12:24.41 & 42:37:11.07 & 5.9 ~ I  & $7.0\pm 2.0 $ & 23.30 & 21.36 &
$21.02^{[2]}$ & 19.20  & d{\it p} \\ L23 &  13:12:25.74 & 42:39:41.5 &
13:12:25.65 & 42:39:38.47 & 3.2 ~  II & $380.0\pm 2.0$ & 23.83 & 23.79
& $>21.72^{[2]}$ & 21.16 & d{\it  d} \\ L24 & 13:12:27.50 & 42:38:00.3
& 13:12:27.49 & 42:38:00.42 & 0.2 ~  I & $22.4\pm 2.0$ & 21.43 & 20.27
& $20.27^{[2]}$ & 18.13 & d{\it  d} \\ S2 & 00:17:40.96 & 15:51:09.3 &
00:17:40.89 & 15:51:11.6 & 2.5 ~ II  & $378.8\pm 29.1$ & - & - & 25.18
&  22.49  &  d \\  S4  &  00:17:41.81  &  15:50:03.4 &  00:17:41.75  &
15:50:03.2 & 1.0 ~ I & $270.2\pm 20.3$ & $23.42^{[5]}$ & $21.84^{[5]}$
 & 20.96 & 18.98 & b \\
S5 & 00:17:43.85  & 15:54:05.0 & 00:17:43.78 & 15:54:04.2 &  1.3 ~ I &
$843.9\pm 51.8$ & $21.68^{[5]}$  & $19.27^{[5]}$ & - & 17.24 & b{\it
p} 
 \\ S9 & 00:17:59.66 &
15:41:59.2 & 00:17:59.59 & 15:41:57.5 &  2.0 ~ I & $5725.\pm 175.$ & -
& -  & 22.15 & 19.19  & b{\it p} \\  S10 & 00:18:03.75  & 15:49:04.0 &
00:18:03.66 & 15:49:04.6 & 1.4 ~ I  & $53.0\pm 12.1$ & 23.08 & 22.90 &
22.23  &  20.87  & d{\it  d}  \\ 
 S11  &  00:18:07.61 &  15:48:27.8  &
00:18:07.63 &  15:48:27.2 & 0.7  ~ I &  $51.4\pm 11.0$ & $19.26^{[5]}$
& $17.72^{[5]}
$ &  - &
16.36 & b \\ S12 & 00:18:18.83 & 15:54:35.4 & 00:18:18.78 & 15:54:35.6
& 0.7  ~ I & $568.2\pm  51.0$ & -  & - & 21.53  & 19.36 & d{\it  p} \\
\hline
\end{tabular}
Notes: -: Not  observed in this passband.  [1]:  Pair with  16V29. 
 [2]:  $I_{HST}$  magnitude from
Windhorst et al. (1995).  [3]: A significant detection on the NIRC but
not on the  LRIS image -- listed $B$ and  $R$ magnitudes are estimated
from  aperture   photometry  at  the  position   of  the  $K^{\prime}$
detection.  [4]: No counterpart visible within 7 arcsec in either LRIS
or  NIRC  image. [5] Estimated from the $B_J$ and $F$
 photometry of Munn et al. (1997)
 *:  Suffix  to  name of  Lynx2  sources  not in  the
`complete' sample of Windhorst et al. (1993).
\end{table}
\twocolumn

The test  selected the simulated counterpart,  rather than
one of the uncorrelated background galaxies, as the best ID for 91 per
cent of the sources. 
\subsection{Magnitudes and Morphologies}
    We examine the NIRC  $K^{\prime}$-band frames and identify the
galaxies  considered the most likely optical counterparts of 
the radio
sources.  Table 1 gives the apparent magnitudes
 as measured using SExtractor, together with RA/Dec co-ordinates of
the optical IDs (from
    our astrometric transforms) and the radio centroids.
Two galaxies, 16V36 and L21, were detected as single
sources in  our analysis of the  LRIS frame, but  appear double-peaked,
and are split  
 into two detections on
the NIRC data. For these we give in Table 1 the
$K^{\prime}$ mag of the two components combined, and consider the two
nuclei separately in  Section 7. For S4, S5 and S11, 
we estimate $B$ and $R$
magnitudes
from the $B_J$ and $R_F$ photometry of Munn et al. (1996), using 
magnitude transforms (Fukugita, Shimasaku and Ichikawa 1995)
 appropriate for ellipticals (on the basis of the
 red colours and bulge profiles of these galaxies).

Table 1 gives the observed 8.44
GHz fluxes  from  Windhorst et
al. (1993, 1995), for Lynx2 and Lilly field sources. 
For the SA68 sources, 8.44 GHz fluxes are estimated
from the 5.0 GHz fluxes of Weistrop et al.  (1987), assuming the
 Condon (1992) SED, for which
$F(8.44)/F(5.0)=0.73(0.76)$ at $z=0(1)$.

Greyscale plots of the identified galaxies will be included in the
 published version (MNRAS).  For each of the optical IDs, we examine the
 LRIS and/or NIRC images by eye, and fit isophotes and radial profiles
to the $K^{\prime}$-band images using {\sevensize IRAF}
`isophote.ellipse'. On this basis we assign 
approximate morphological classifications (Table 1).

Most  of  the galaxies  have approximately  exponential
 (disk) profiles and are classed here as (i) apparently
`normal' disks (`d') - isolated,
 symmetric, with single nuclei, (ii) disturbed disks (`d{\it d}')
 with visible asymmetry or isophotal  twist, or (iii) disk galaxies in
 close pairs or groups or with double nuclei (`d{\it p}'), which may
 be interacting or merging. Of course, some galaxies in class `d' may
 have interaction features not visible in these ground-based images. 
In addition, there are QSOs seen as high surface brightness point sources
 (`Q'), one QSO  within an apparently
 merging system (`Q{\it p}'), bulge galaxies (`b'), i.e.
with profiles closer to  the form ${\rm
 exp}~[-r^{0.25}]$ than ${\rm exp}~[-r]$,  and bulge galaxies in close
 pairs (`b{\it p}'). Our classifications of the Lilly field sources
 are in general agreement with Windhorst et al. (1995).
\section{Spectroscopy: Results}
On the  Lynx2 multi-slit  mask, 12 slits  were dedicated to  16V radio
sources.  Spectra were successfully extracted for 16V10*, 13, 15*, 17,
25, 31, 34 and  36.  We failed to obtain spectra for  16V21, 22 and 35
due to inaccurate  slit positions -- the slits  had been positioned on
the VLA  co-ordinates, which were a  few arcsec offset  from the faint
optical counterparts. In the case of 16V08, the slit position appeared
correct but the galaxy was too faint ($R=24.53$) to see any significant
spectrum.   The faintest successfully  observed galaxy
was 16V10*, with $R=24.27$.  The spectroscopic slit covered only one of
the two nuclei of the  merging 16V36, while for the paired 16V15*
we obtained spectra for both the radio ID
and its companion (hereafter 16V15B).

 An additional long-slit observation provided a spectrum for
the previously missed 16V22, and also 16V24, but the latter proved too
faint ($R=24.15$) to identify  any spectral features.  The Lilly field
multi-slit  mask provided  good  spectra for  L4,  6, 9,  18, 21,  and
24. Objects  L10,  13  and  16  were missed  due  to  inaccurate  slit
positions   (a  discrepancy  between   VLA  and   optical  astrometric
solutions), while  L5 ($R=24.18$) and L23 ($R=23.79$)  were too faint.
Long-slit  spectroscopy of SA68  sources provided  spectra for  S4 and
S10. S2 was also observed but, with $I=25.18$, was too faint.

We  have  spectroscopically observed  22  of  the  50 radio  IDs,  and
 obtained `good' spectra for 17,  the remaining 5 being too faint. All
 except for  16V10*, 16V15* and L21 belong  to `complete' radio-source
 lists.  The  spectra  were  examined  using an  enhanced  version  of
 {\sevensize  IRAF}  `splot',  to  identify significant  emission  and
 absorption features, and measure the
 equivalent  widths   (hereafter  EWs) and  
 FWHM (from a Gaussian fit) of emission lines. Redshifts 
were determined from the one or two strongest emission lines. 
Table 2 gives the EWs (errors are typically 0.5--$1.0\rm \AA$; see 
Section 8).

We  measure redshifts from  the 17  `good'  spectra, and
consider these to  be reliable, as all are  based on multiple spectral
features.   In addition,  Windhorst et  al. (1995)  give spectroscopic
redshifts of $z=0.322$  for L3, $z=0.302$ for L13,  $z=0.180$ for L15,
$z=2.561$ for  the QSO L17, and  $z=0.401$ for L19,  Weistrop et
al. (1987)
give  $z=0.168$  for S11 and Munn
et al. (1997)  $z=0.3493$ for S5, $z=0.0531$ for  16V28 and $z=0.0535$
for 16V29.   Thus in total we have  a subsample of 26  radio sources with
known redshifts, hereafter the `redshift sample'.
 
The redshift sample cannot be described as purely radio-selected, as
there is  an  additional  optical  selection  bias, approximately  in  the
$R$-band (about
the centre  of the spectrometer  wavelength range). The  magnitudes of
the radio IDs extend faintwards of $R\sim 25$, while all the redshift
sample galaxies have   $R<24.3$, and all the
  redshifts from  previous  data are for 
$R<20.5$ galaxies.
 Figure 2 illustrates the optical bias of the redshift
sample by comparing its $R$
magnitude distribution with that of all the radio IDs.

Of the 17 radio IDs with LRIS spectra, 14 are emission-line galaxies 
and 3 are
QSOs.  Emission  features  typically  observed are  $\rm  H\beta$  and
$\rm[OIII]5007\AA$,  together  with $\rm  [OIII]4959\AA$,  and at  the
higher redshifts, $\rm [OII]3727\rm \AA$.  Some galaxies show
Balmer lines  $\rm H\gamma$ (in either  emission or
absorption), and $\rm H\delta$, 
and the broader absorption features $\rm H 3969\AA$ and
$\rm K 3934\AA$. Only 16V10* appears to show $H\delta$ in emission and
the reality of the line is is uncertain as it is only 2--$3\sigma$
and, although its wavelength is
consistent with the redshift
from the other features, it is
also within $\sim 3\rm\AA$ of a sky line. 
The  [OII]3727 line  of 16V10*  appears on  the 1D  plot to  be `lost'
 amongst sky lines, but the 2D spectrum reveals it to be very different from
 the sky features and a real emission line. 

The  companion galaxy  16V15B shows  only absorption  lines, including
$\rm G 4304\AA$  and $\rm H\gamma$ (EW $\sim  1.0\rm \AA$), the latter
giving $z=0.4855$ with a line-of-sight velocity only $71\pm 33$ km $\rm
s^{-1}$ greater  than 16V15. Another pair, 16V22  and 16V34 are  also
at closely similar redshifts, with 
$\Delta(v)=35\pm17$ km  $\rm  s^{-1}$,  but are  separated  by 180  arcsec
($\simeq 1.39$ Mpc) on the sky. 

The
QSO spectra may show $\rm MgII 2796.4, 2803.5\AA$ (as broad emission and/or
narrow absorption lines) and/or narrow $\rm [NeIV]2424,2426\AA$ or $\rm OIII
3341\AA$ emission.

\begin{table}
\caption{Redshifts (errors  $\pm 0.0001$) and rest-frame  EWs (in $\rm
\AA$,  emission   positive,  absorption  negative)   of  significantly
detected  spectral  lines,  for  galaxies  (above)  and  QSOs  (below)
identified as radio sources.}
\begin{tabular}{lcccccc}
\multispan{7} Galaxies \\ \hline Source  & $z$ & [OII] & $\rm H\delta$
& $\rm H\gamma$ & $\rm H\beta$ & [OIII] \\
\smallskip
       &     & 3727 &    &    &    &  5007 \\ 
16V10* & 1.0838  &    7.37 &  2.83?  &  - & - & - \\
16V15* &  0.4854 &   -  &  - &    1.33 &   9.59 &   6.16 \\
16V22 & 0.4259  &    - &  -   & 5.19 & 8.92 & 30.19 \\ 
16V25 &  0.7273  &    8.78 &  -1.94 & $<$ & $<$ & $<$  \\
16V31 & 0.4977  &    - &  -8.14  & -2.14 &   3.09 &   6.29 \\
16V34 & 0.4257  &    - &   - &   $<$   & 4.54  &  3.37 \\
16V36 & 0.4073  &    - &  - &   $<$ & 4.82 &   2.28 \\
L6 & 0.4946  &   -  & - &  - &   7.83  &  3.81 \\
L9 & 0.6986  &   15.45 &  -4.44 &  $<$ &    8.61  &  6.74 \\
L18 & 0.7648  &    5.68 &  -5.22  & $<$ & $<$ & $<$ \\
L21 & 0.2905  &    - & - &  - &     12.30 &  40.19 \\
L24 &  0.3155  &    - &   -  &  -  &  4.05  &  2.70 \\
S4  & 0.7125 & 7.54 & -2.51 & $<$ & $<$ & $<$ \\
S10 & 0.9923  & 52.50   & $<$  &  -  &   - & -  \\
\hline
\end{tabular}
\begin{tabular}{lcccc}
\multispan{5} QSOs \\
\hline
Source &  $z$ & [NeIV] & MgII & OIII \\
\smallskip
      &          & 2424/6 & 2796/2804 &  3341 \\
16V13 & 1.8330 & - &  -18.68, -16.08 &  \\
16V17 & 1.6550 & - & - & 0.76 \\ 
L4    & 2.5580 & 0.54, 0.39 &  - & - \\
\hline
\end{tabular}

(-: line is outside observed $\lambda$ range. $<$: line is not detectable
at $2\sigma$ level). 
 \end{table}

\begin{figure} 
\psfig{figure=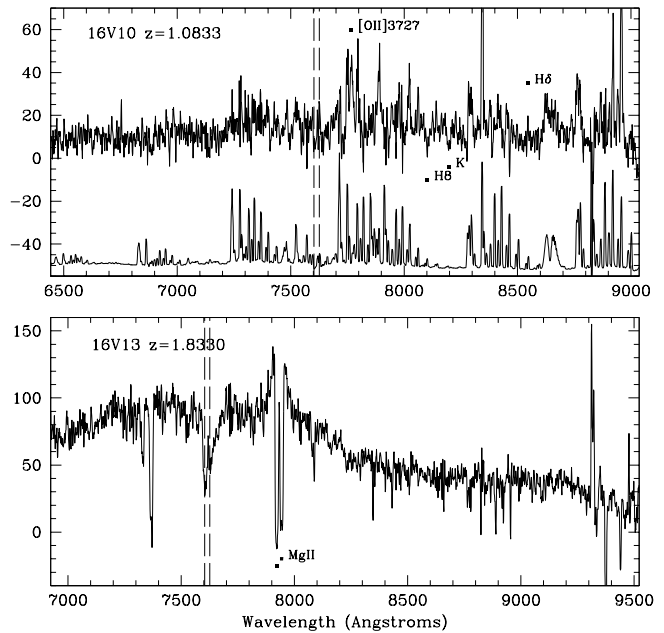,width=90mm}
\caption{Spectra  of the  17 radio  sources and  the  companion galaxy
16V15B.  Significantly detected lines are marked, above (emission) and
below   (absorption)   the    spectra.   The x-axis is observer-frame 
observer-frame, and  the y-axis scale is counts  $\rm pixel^{-1}$ (not
flux calibrated).  The spectrum of the sky background (which can
produce spurious  features in the galaxy spectra)  is shown underneath
16V10*.  Vertical dashed lines  show the wavelengths of
sky absorption features.}
\end{figure}

\begin{figure}
\psfig{file=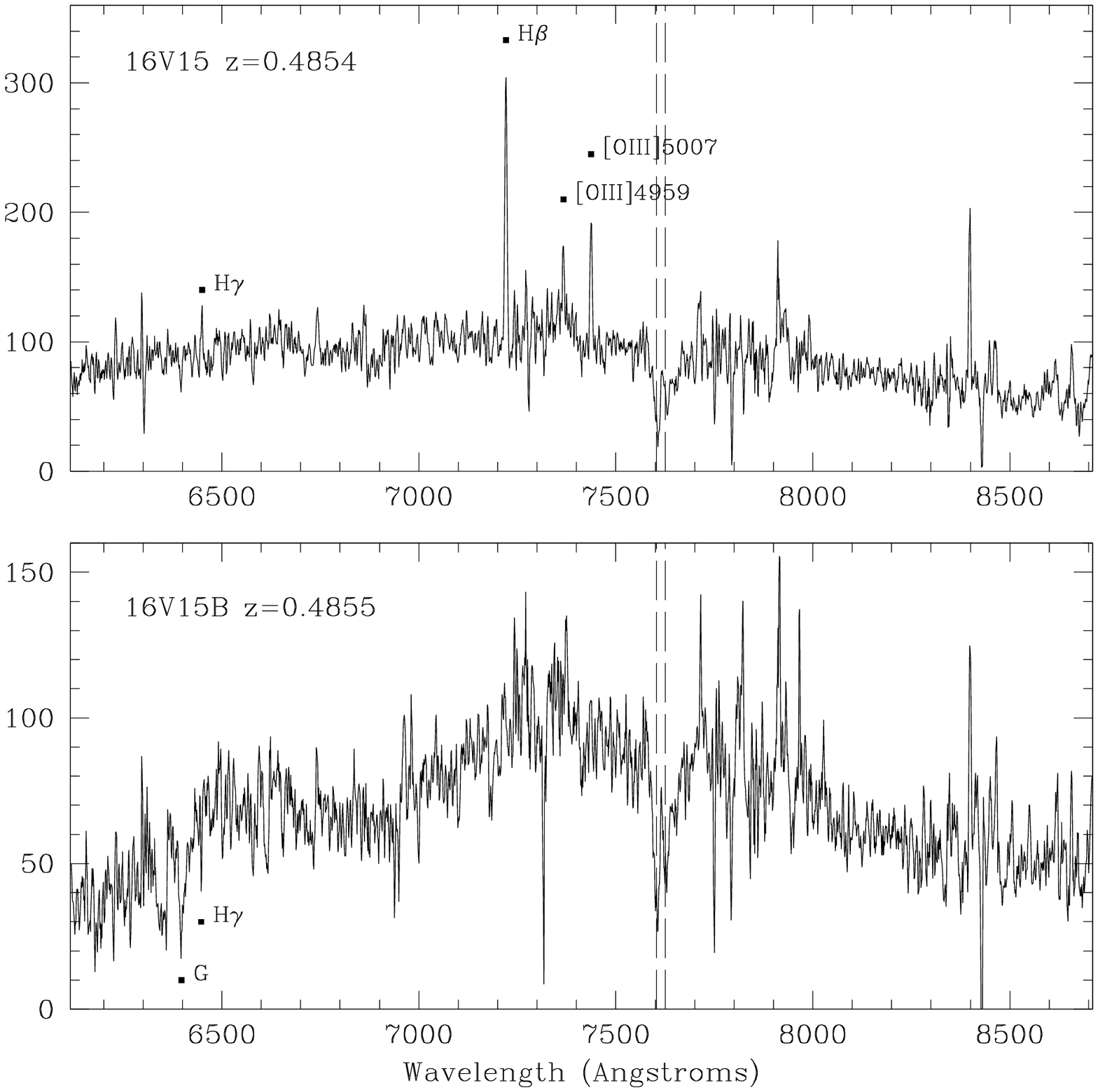,width=90mm}
\end{figure}

\begin{figure}
\psfig{file=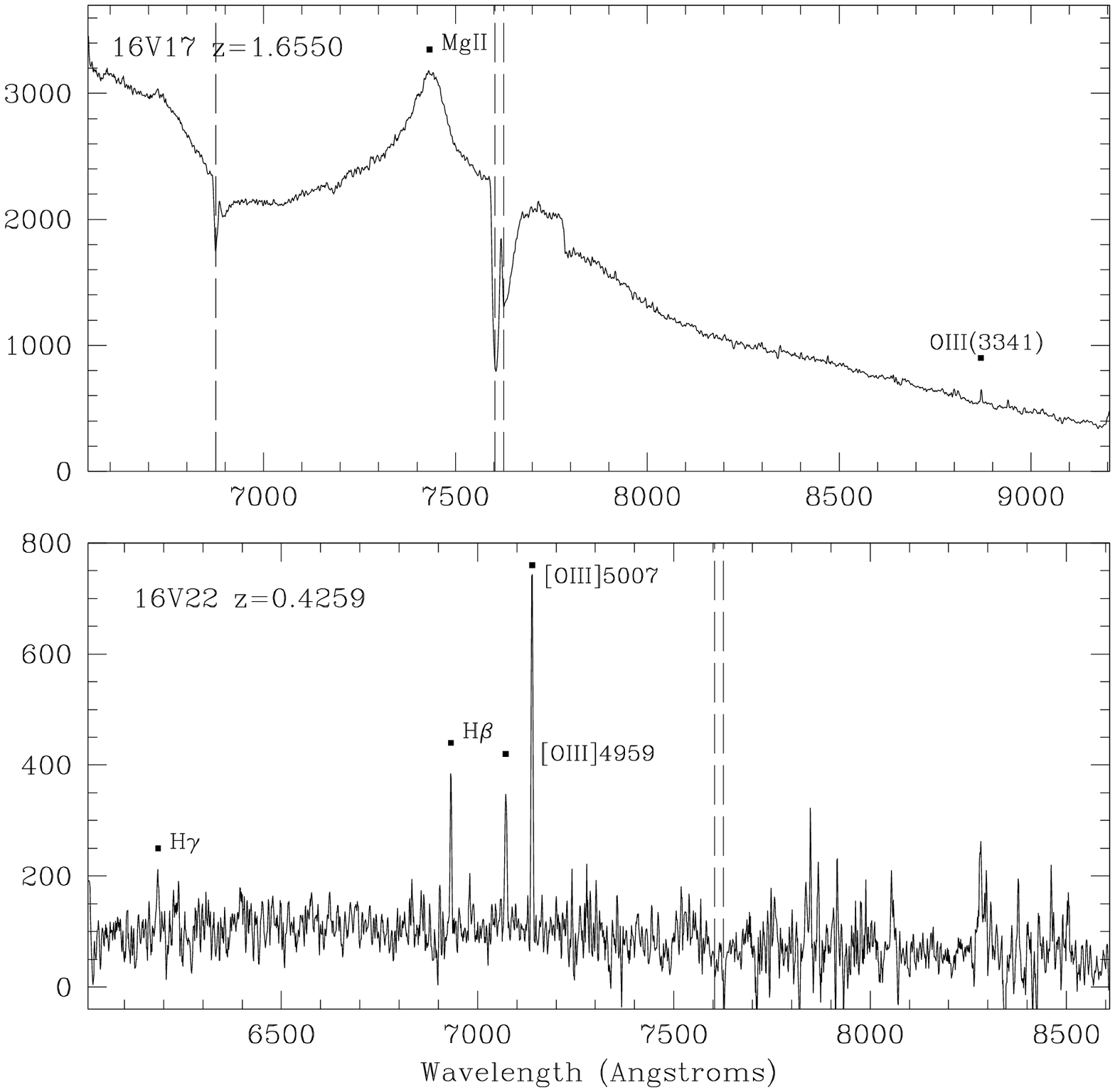,width=90mm}
\end{figure}

\begin{figure}
\psfig{file=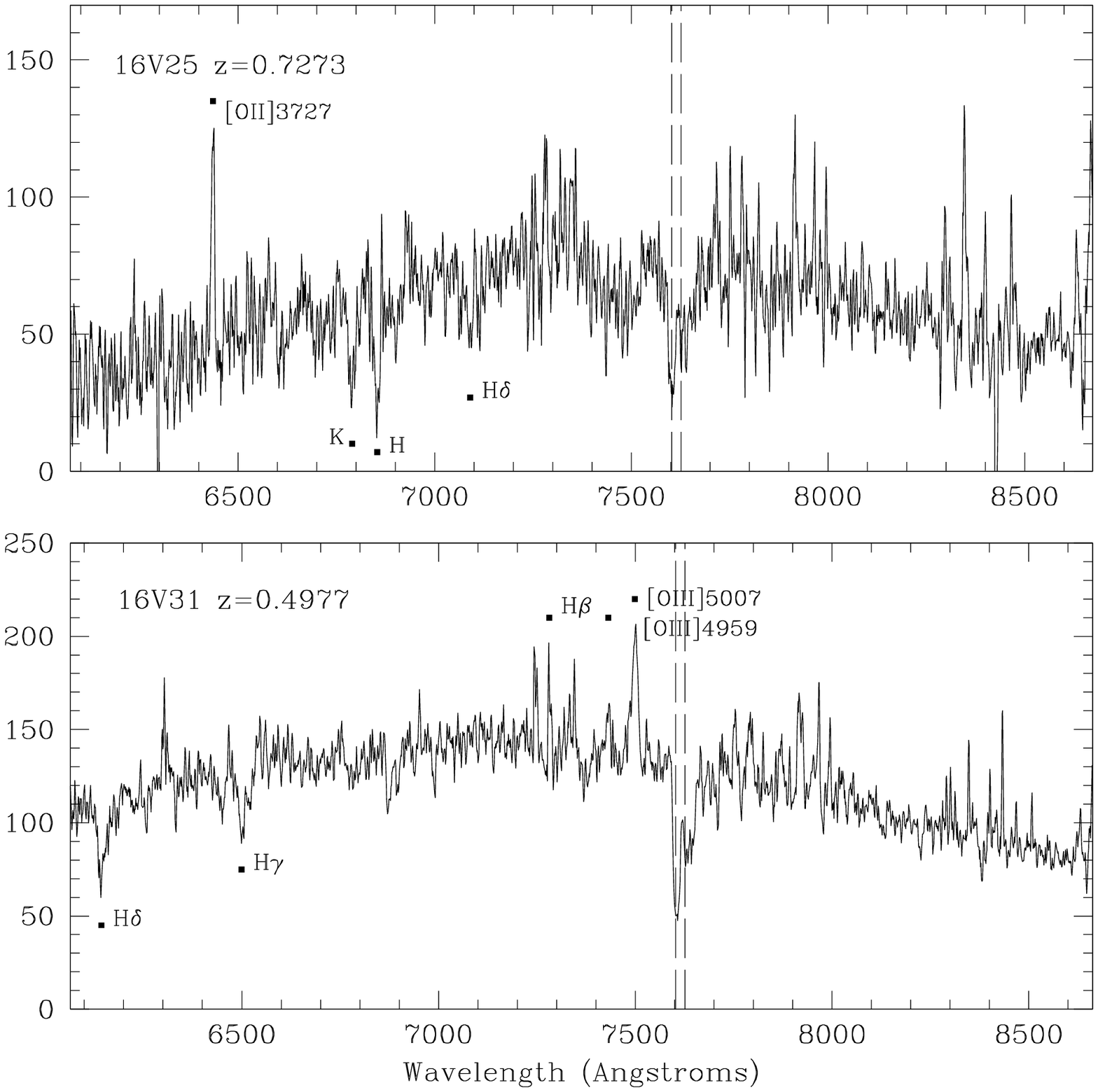,width=90mm}
\end{figure}

\begin{figure}
\psfig{file=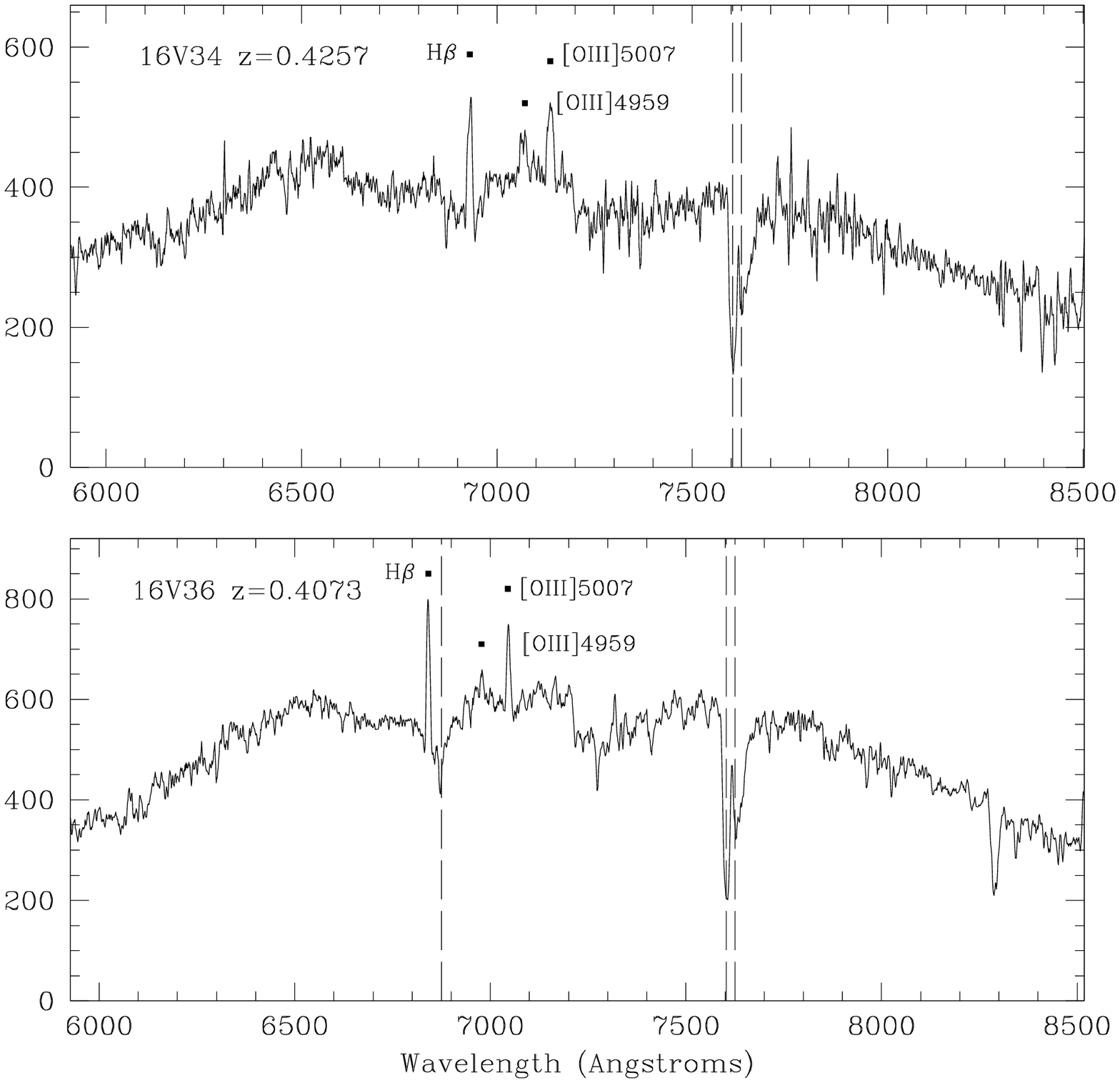,width=90mm}
\end{figure}

\begin{figure}
\psfig{file=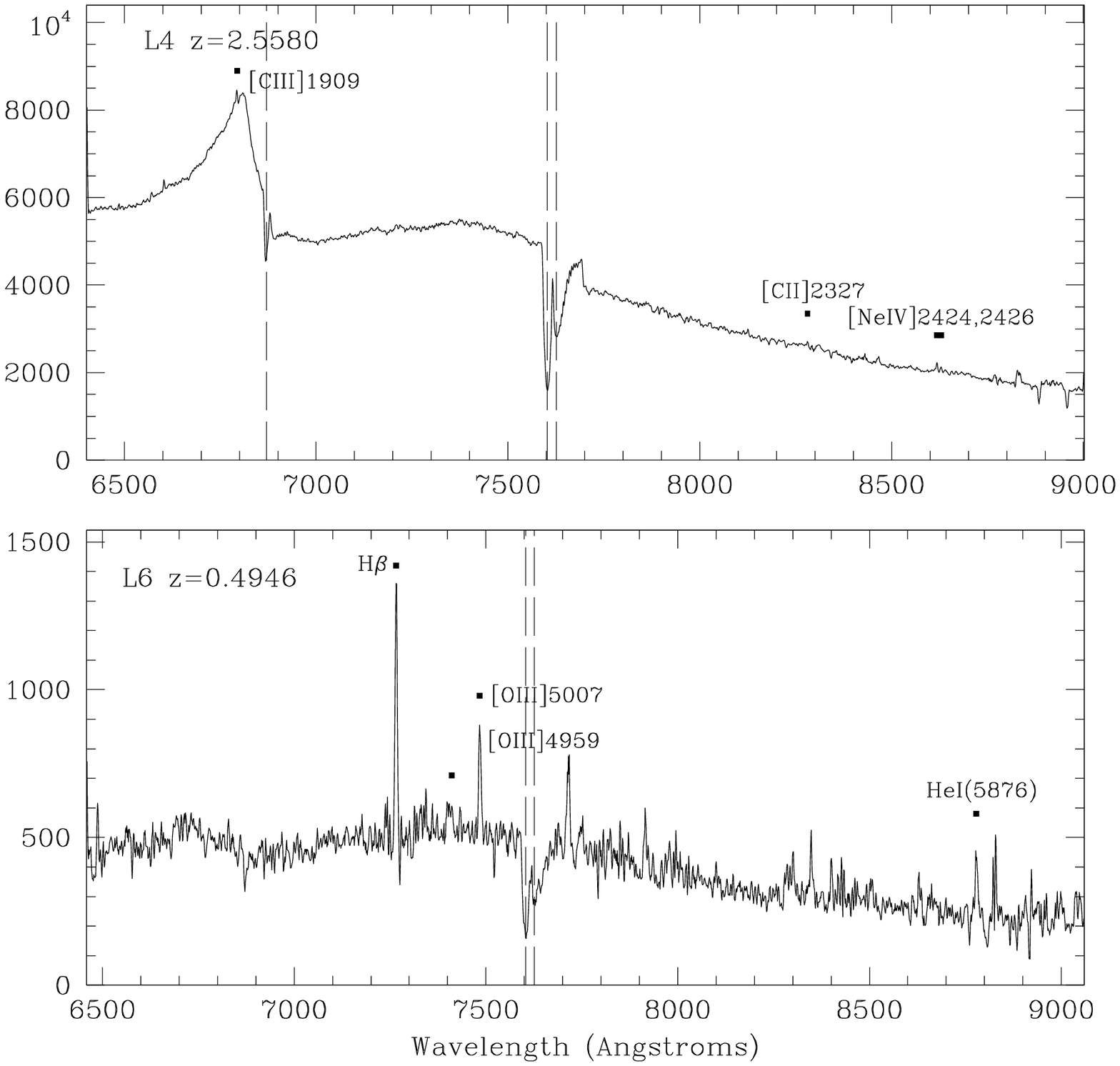,width=90mm}
\end{figure}

\begin{figure}
\psfig{file=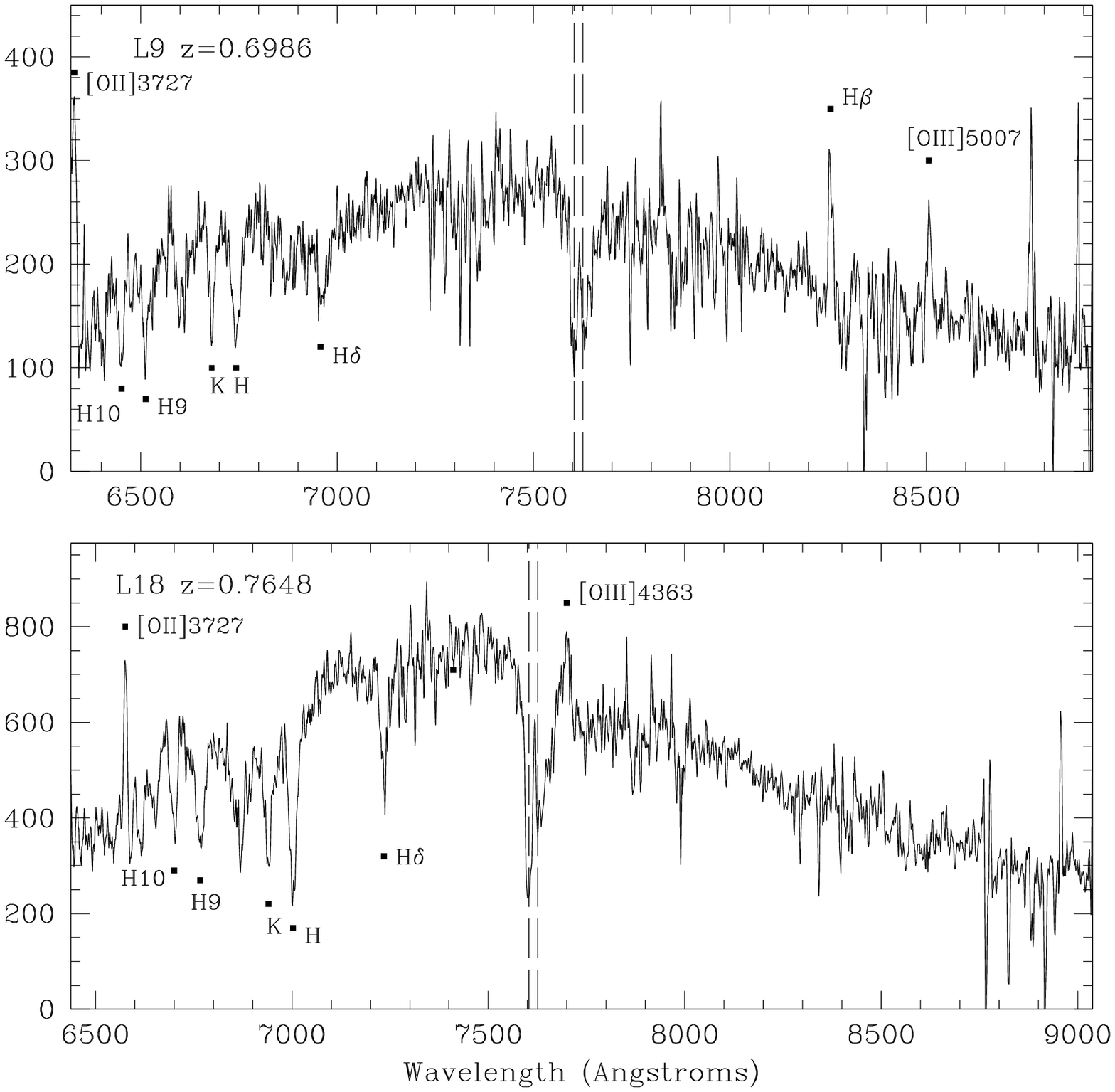,width=90mm}
\end{figure}

\begin{figure}
\psfig{file=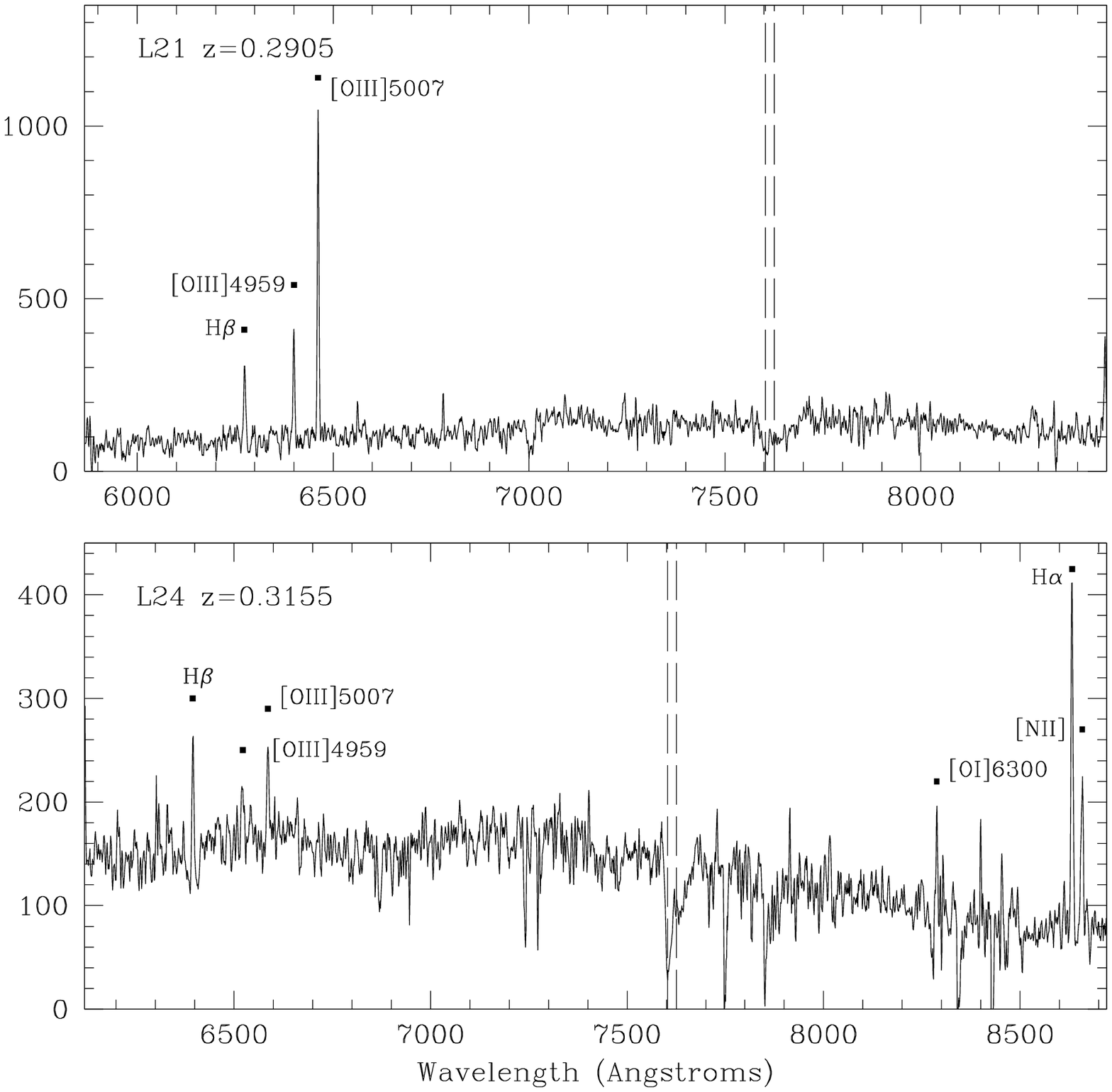,width=90mm}
\end{figure}

\begin{figure}
\psfig{file=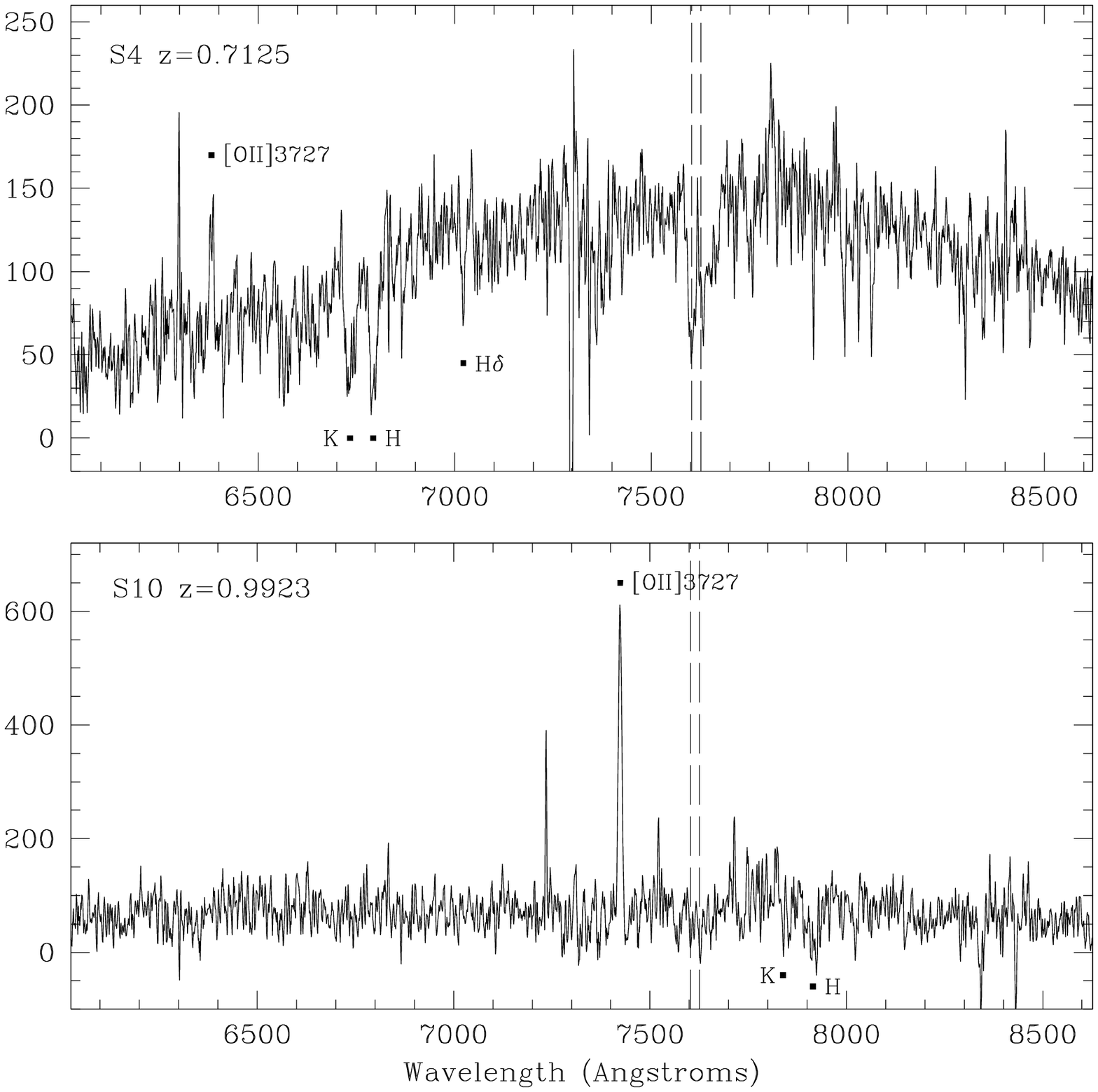,width=90mm}
\end{figure}
\begin{figure}
\psfig{figure=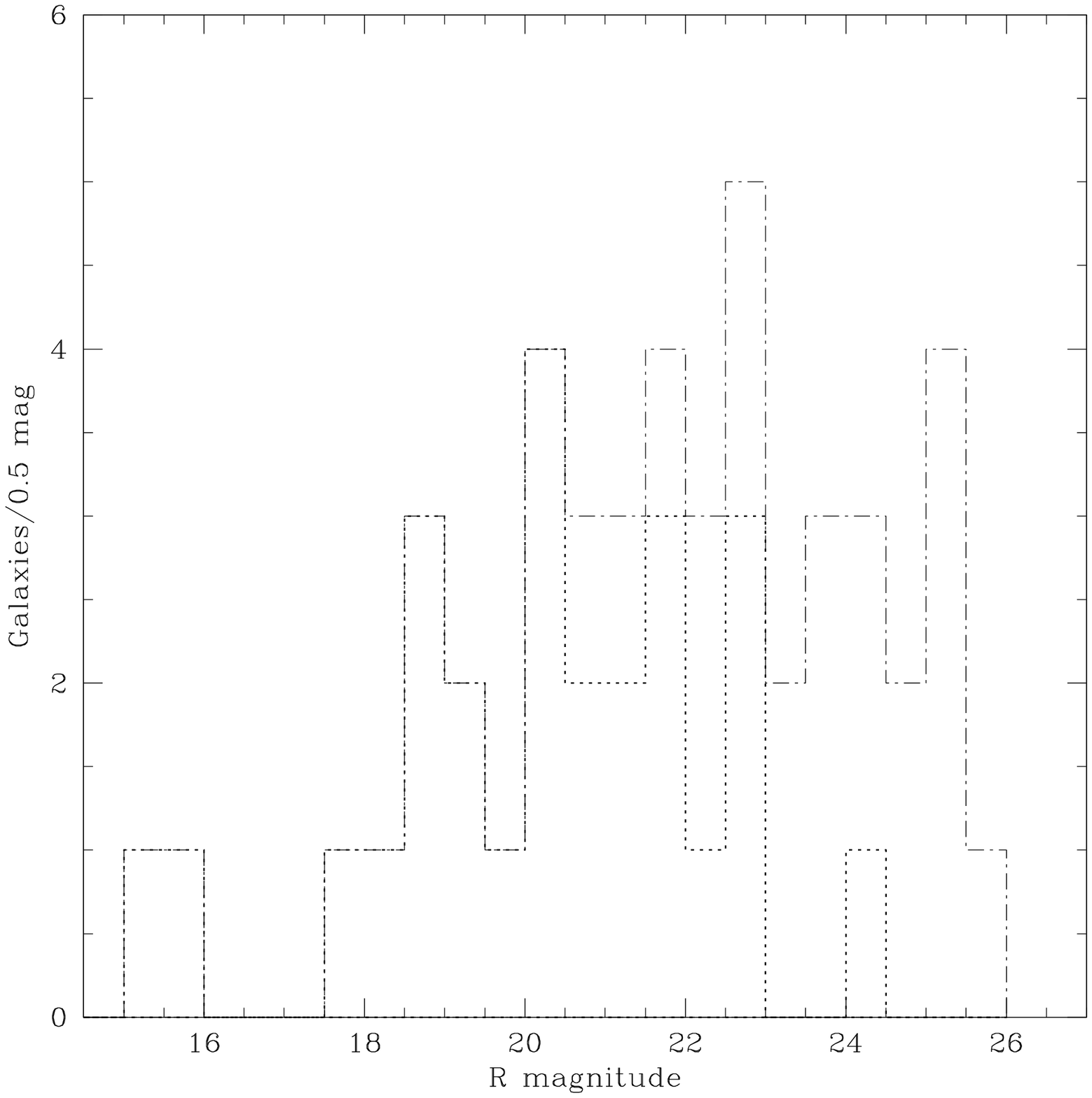,width=90mm}
\caption{The $R$ magnitude distributions of the radio-source IDs in
Table 1 (we have $R$ magnitudes for 47/50) (dashed histogram),
and the subset for which we have spectroscopic redshifts  
(we have $R$ magnitudes for all 26) (bold dotted histogram).}
\end{figure}
\section{Colours and Luminosities}
\subsection{Optical Flux}
\begin{figure}
\psfig{file=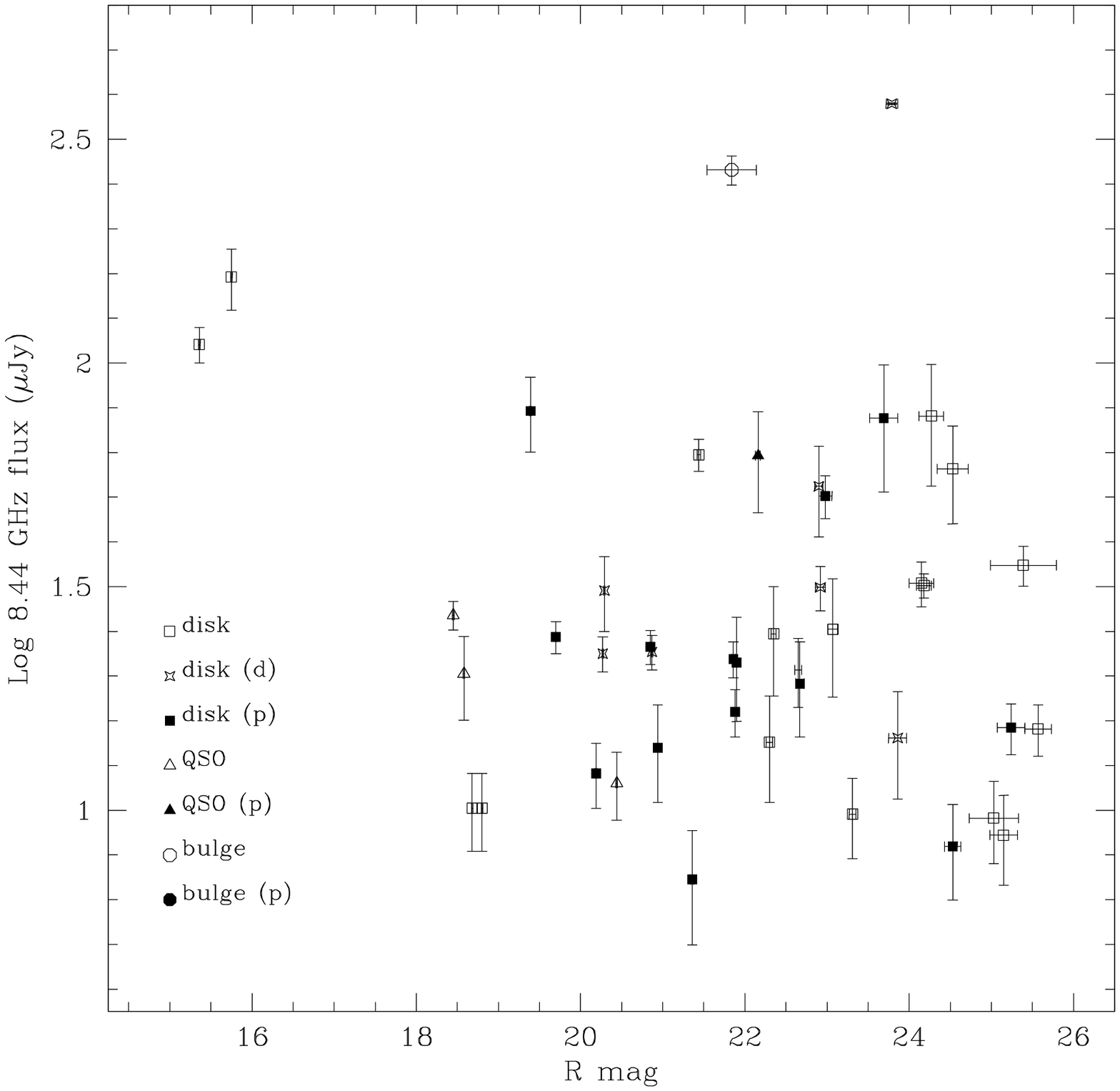,width=90mm}
\caption{$R$ magnitude against 8.44 GHz flux for 44 radio-selected galaxies
(symbol types showing our morphological classifications).}
\end{figure}
We have  identified likely optical counterparts for  51 radio sources,
and of  these 44 are observed  in the $R$-band.  Figure  3 shows their
$R$ magnitude against 8.44 GHz flux, with error bars.

An important  question is whether  radio-selected galaxies tend  to be
bluer (e.g. from star-formation) or redder (e.g. from dust)
 than other galaxies at
similar apparent magnitudes. Figure 4a compares the $B-R$
distributions of the 47 radio source IDs with $B$ and $R$ magnitudes,
with that  of all  519 $R<24.0$ galaxies
(including  the radio IDs)  on  the combined  Lynx2 and  Lilly
fields.  The radio-selected sample contains a higher proportion of
$B-R\sim  1$ galaxies,  but the mean  $B-R$ is only slightly redder, 
0.71 compared to  0.62; only 3/47 of the radio source IDs are very
red  ($B-R>2$),  while  the proportion  (16/44) 
of  very blue ($B-R<0.2$) galaxies is similar to
the $R<24.0$ sample. Figure 4b compares the 
$B-I$ distributions of the 24 radio IDs
observed in $B$ and $I$ with that of all 
$R<24.0$ galaxies on Lynx2 - these are again very similar.

We conclude from this the radio detections generally have similar
colours to optically-selected star-forming galaxies, and hence 
if they are
similar types of galaxy they
have little  ($\Delta(B-R)<0.5$  mag) or  no additional
reddening.
\begin{figure}
\psfig{file=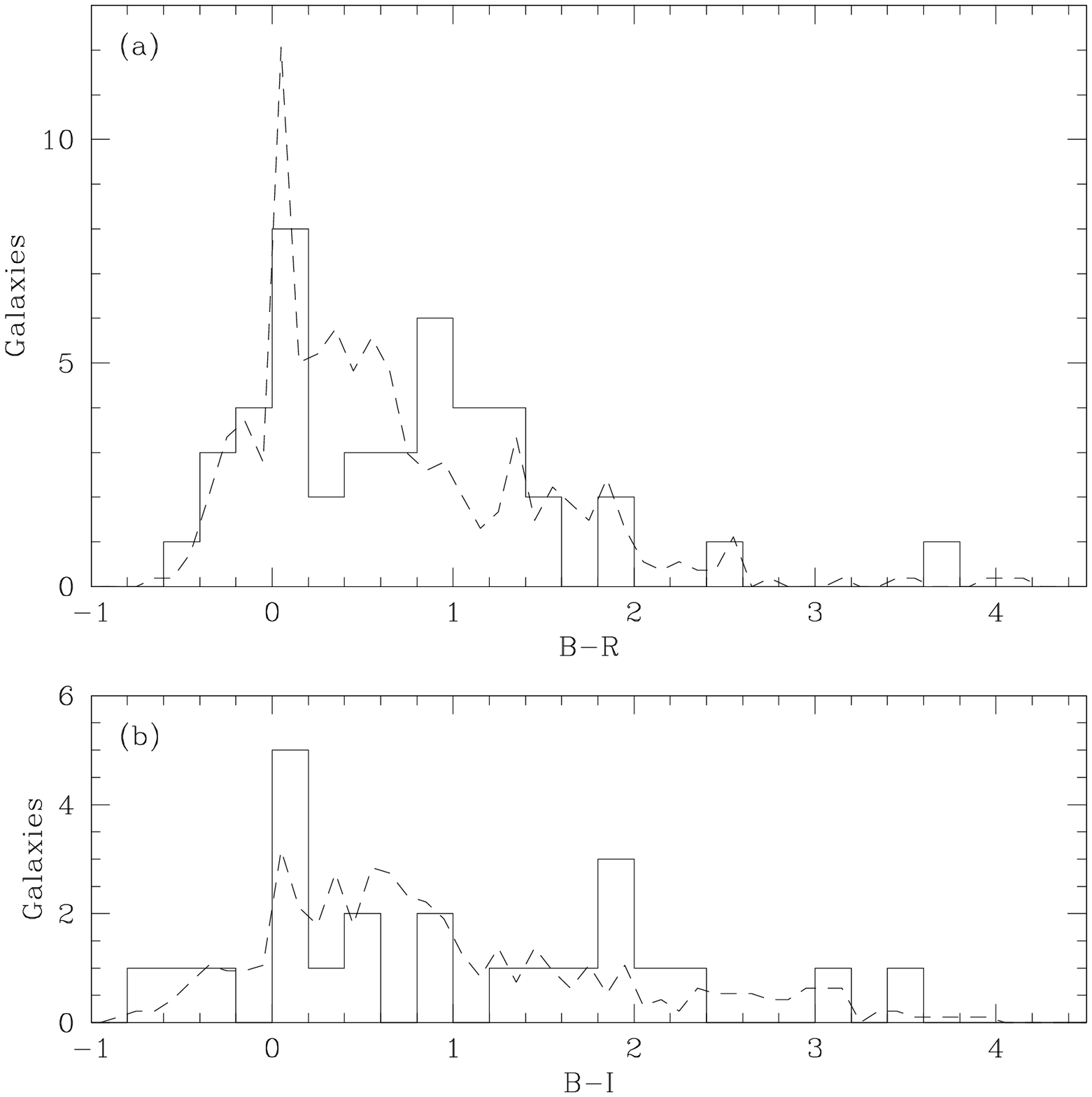,width=90mm}
\caption{Histograms of  (a) the $B-R$  and (b) $B-I$  distributions of
the galaxies identified  as radio sources, compared with  those of all
$R<24.5$ galaxies  (dashed line)  on the Lynx2  and (for  $B-R$) Lilly
fields, normalized to the same total number.}
\end{figure}

The colours of the redshift sample can be interpreted more
directly  by comparing  with  a grid  of models. We use models
 (from  Roche et
al. 2001a)
computed with the `Pegase 2' package (Fioc and Rocca-Volmerange 1997,
1999) to represent normal (non-starburst)  spirals from Sa to Im, and
E/S0 galaxies. 
We assume the
time-redshift  relation of an  $\Omega_m=0.3$, $\Omega_{\Lambda}=0.7$,
$H_0=55$ km $\rm s^{-1}Mpc^{-1}$ Universe, with model galaxies 
formimg 16 Gyr ago, at $z=6.05$.  The E/S0 galaxies evolve passively
(no star-fomation) at $z<2$. The bluest starburst galaxies are
represented as a  constant-SFR model observed at an age of  50 Myr,
 with  no  dust  reddening,  which is  redshifted
without evolving. The models are computed with a stellar initial
mass function (IMF) following the Salpeter slope, $x=2.35$ where ${d{\rm
N}\over d  {\rm M}}\propto \rm M^{-x}$, at  $\rm 0.7<M<120 M_{\odot}$,
but flattening  to $x=1.3$  at $\rm 0.1<M<0.7  M_{\odot}$ (see  
Holtzman et al. 1998).

Figure  5 shown the redshift  sample on  a $B-R$
vs. $z$ plot, on which  the majority of the galaxies are close to the spiral
models;  L3 and  16V22 are  significantly bluer  suggesting unreddened
starbursts, L21, L9, 16V25 and possibly 16V10* are up to 1 mag redder,
 and the elliptical S5 is consistent with
 the passive E/S0 models. 
Figure 6 shows $R-K$ vs. $z$ -- 
again, most of the galaxies are consistent with the spiral models;
 L3 and 16V22 are bluer, while 16V25, 16V10*, L9, and also 16V31, are
 again redder. 
 
\begin{figure}
\psfig{file=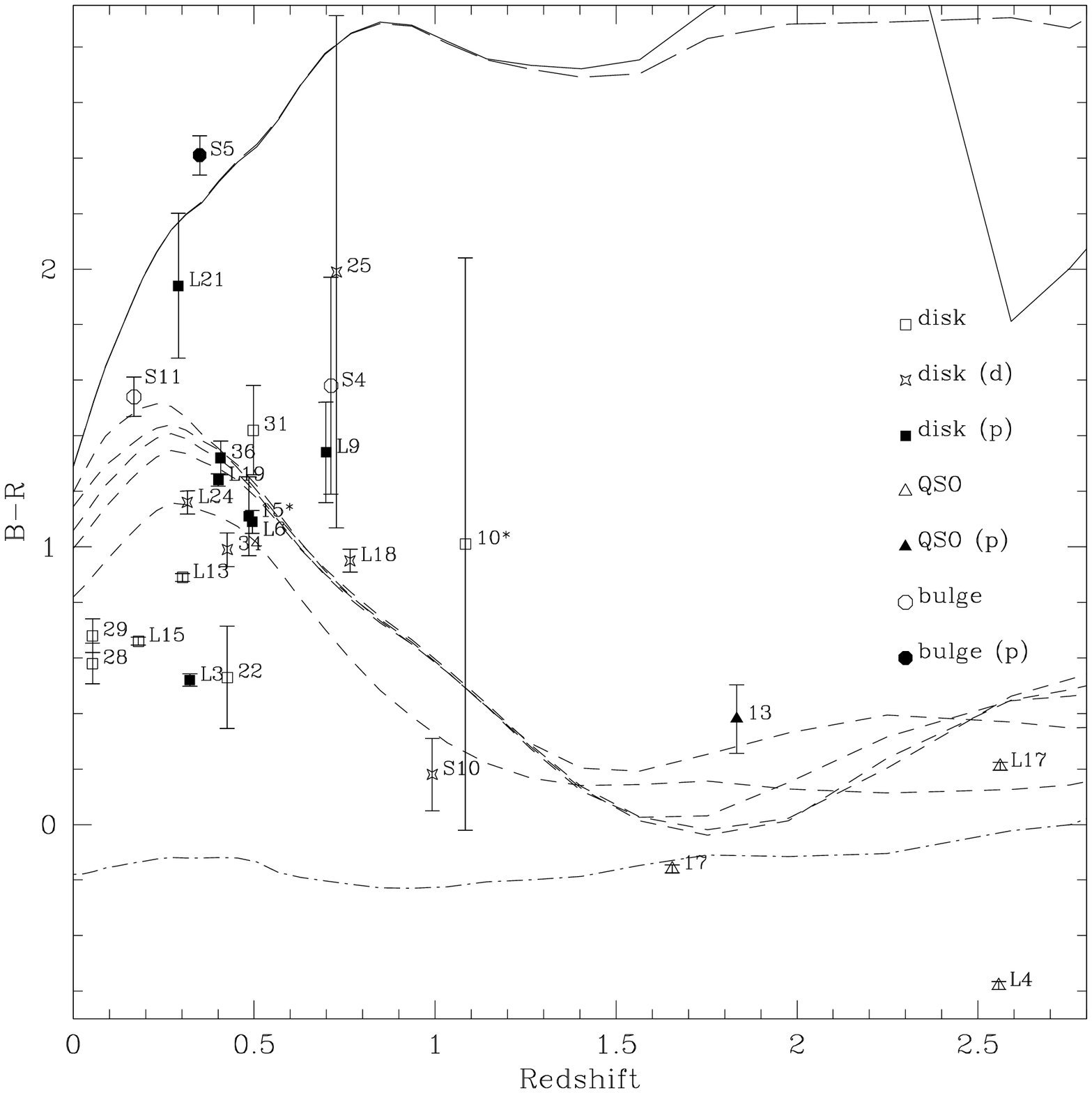,width=90mm}
\caption{$B-R$ colour (AB  system) against redshift for radio-detected
galaxies (symbols indicating types),  compared to models for
evolving  spirals --  Sa,  Sb,  Scd, Sc  and  Im (short-dashed  lines,
reddest  to   bluest),  early-type  galaxies  --  E   (solid)  and  S0
(long-dashed), and a  50 Myr starburst (dot-dash).  The  labels of the
Lynx2 galaxies omit the `16V' prefix.}
\end{figure}

\begin{figure}
\psfig{file=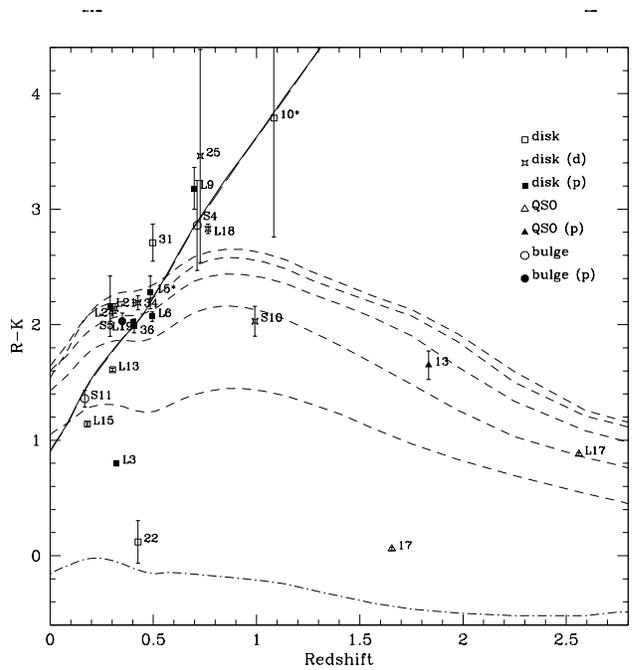,width=90mm}
\caption{$R-K$  (AB   system)  against  redshift   for  radio-detected
galaxies, with models as Figure 5} 
\end{figure}
\subsection{Radio vs. Optical Flux}
Condon  (1992) models  the radio  SED of  a star-forming  galaxy  as a
summation  of thermal  ($\nu^{-0.1}$)  and synchrotron  ($\nu^{-0.8}$)
components,
$$F_{\nu}(\nu)=f_{>5}~{\rm  SFR}[5.3\times 10^{28}\nu^{-0.8}+5.5\times
10^{27} \nu^{-0.1}]$$ where $F_{\nu}$ is in ergs $\rm s^{-1} Hz^{-1}$,
$\nu$ is  frequency in  GHz and  $f_{>5}$ is the  fraction by  mass of
stars  formed   with  $M>5M_{\odot}$.   For  the  IMF   adopted  here,
$f_{>5}=0.3095$ (compared to $f_{>5}=0.1863$ for the standard Salpeter
IMF).  In  Roche et al. (2001a)  this  relation was  used  to  predict the 
radio/optical flux ratios for two sets of
galaxy models, representing (i) normal, steadily evolving
galaxies of Hubble types Sa-Im  (colours plotted on Figures
5 and 6), and (ii) starbursting galaxies where a
starburst (50 Myr age constant SFR model with $A_{V}=0.7$ mag
reddening) model  SED is  added to each  normal galaxy,  normalized so
that the 
total SFR and hence radio luminosity is increased by $\sim 0.8$ dex. 
  For E/S0 galaxies we adopted an
observational radio/optical ratio, from the
$L_{rad}\propto L_{B}^{2.2}$ relation  of Sadler, Jenkins and Kotanyi
(1989).

 Figure 7  shows the  8.44 GHz to  $R$-band flux ratio  vs.  redshift,
compared with the Roche et al. (2001a)  models for normal and starburst
galaxies. The elliptical
 flux ratio is plotted for $M_B=-21.5$, and will be 0.48 dex
higher for each 1 mag brighter in optical luminosity.Also shown is a
 model for Seyfert 2s based on the tabulated
mean SED of Schmitt et al. (1997).
Figure 8 shows the same for the 8.44 GHz to $K^{\prime}$ ratio.
Both figures show that: (i)
 four $z<0.4$ disk galaxies and  the elliptical S11 have relatively low
ratios of radio to $R$ or $K^{\prime}$ flux, consistent with normal
non-starburst spirals and ellipticals; (ii) three of the QSOs have similar
low ratios
but 16V13 is an order of magnitude higher; (iii)
 most radio IDs 
have ratios in the range of our starburst
models, but (iv) seven (16V10*, 22, 25, 13, S4, S5 and S10) are
significantly higher 
-- implying extreme  SFRs, greater dust extinction,
and/or radio-loud AGN.
\begin{figure}
\psfig{file=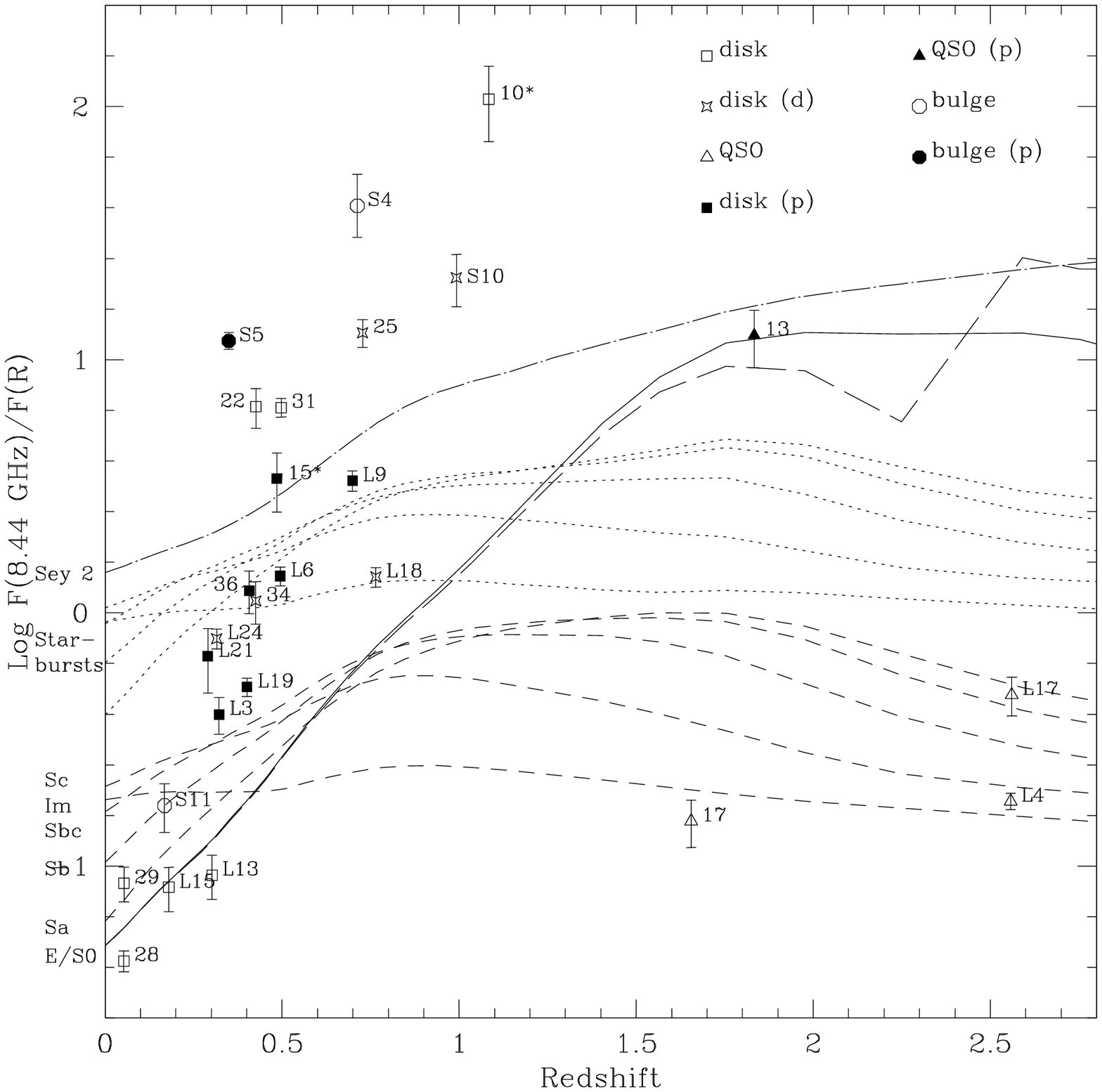,width=90mm}
\caption{Ratio  of  8.44  GHz  to $R$-band  flux  ($F_{\nu}$)  against
redshift,   for  the   radio-detected  galaxies,   symbols  indicating
types.  These observations  are compared  with evolving  galaxy models
 for spirals (dashed, models labelled on axis), starburst
spirals (dotted, in same order as normal spirals), Seyfert 2 (dot-long
dash), and E/S0s (solid/long-dashed) at $M_B=-21.5$.}
\end{figure}

 \begin{figure} \psfig{file=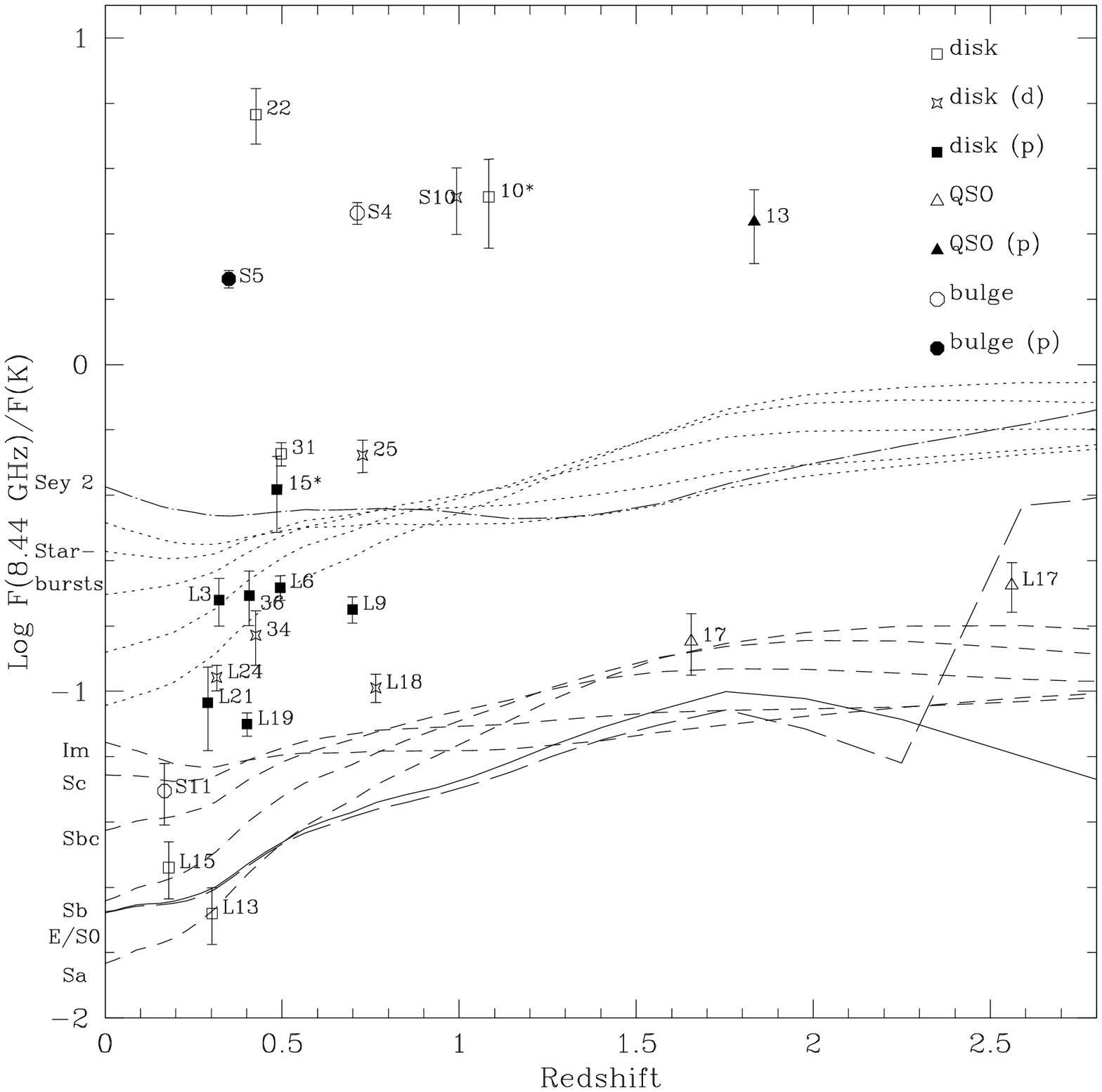,width=90mm}
\caption{Ratio of 8.44 GHz to $K^{\prime}$ band flux ($F_{\nu}$)
against 
redshift, for the radio-detected galaxies observed in $K^{\prime}$, compared with the same models as plotted in Figure 7.} 
\end{figure}

\subsection {Radio and Optical Luminosity}
Radio luminosities and $B$ and $K^{\prime}$ absolute magnitudes 
are estimated from the observed fluxes
and magnitudes as described in Appendix 1. For S5 and S11, we make use
of $B$ and $R$ magnitudes derived from   Munn et
al. (1997) (Table 1) in deriving the $M_B$ and $M_K$, but for
 S4 use only  our own $I$ and $K^{\prime}$
photometry
($I$ being close to rest-frame $B$ at this redshift).

\begin {table}
\caption{Estimated absolute magnitudes $M_B$ and $M_K$ in rest frame
$B$ and $K^{\prime}$ respectively, and 8.44 GHz restframe radio
luminosities $L_{8.44}$ for the 26 galaxies in the  redshift sample.}
\begin{tabular}{lcccc}
\hline
\smallskip
Source &  $M_B$ & $M_K$ & ${\rm log}~L_{8.44}$ & Redshift \\
16V10* & -22.37 & -24.69 & 40.76 & 1.0838 \\
16V13 &  -25.61 & -26.18  & 41.17 & 1.8330 \\
16V15* & -21.02 & -23.40 &  39.43 & 0.4854 \\
16V17 & -27.36 & -27.50 &  40.59  & 1.6550 \\
16V22 &  -19.87 & -19.86 &  39.29 & 0.4259 \\
16V25 &  -22.08 & -24.71 &  39.99 & 0.7273 \\
16V27 & [1] & -30.69 &  42.36 & 1.680 \\
16V28 & -21.71 & $-23.56^{[1]}$ &  38.08 & 0.0531 \\
16V29 &  -21.28 & $-23.09^{[1]}$ &  38.23 & 0.0535  \\
16V31 & -21.65 & -24.37  & 39.92 & 0.4977 \\
16V34 & -22.19 & -24.53 &  39.47 & 0.4257 \\
16V36 &  -22.91 & -25.08 &  39.82 & 0.4073 \\
L3 &  -21.51 & -22.28  & 38.79 & 0.322 \\
L4 & -28.57 & $-28.23^{[1]}$  & 41.11 & 2.5580 \\
L6 &   -22.14 & -23.78 &  39.44 & 0.4946  \\
L9 &  -22.77 & -25.15  & 39.80 & 0.6986 \\
L13 & -22.67 & -24.38  & 38.65 & 0.302 \\
L15 & -21.42 & -22.75  & 38.16 & 0.180 \\
L17 & -27.50 & -27.63 & 40.73 & 2.561 \\
L18 &   -24.00 & -26.17 &  39.90 & 0.7648 \\
L19 & -22.57   & -24.76 & 39.91 & 0.401 \\
L21 &  -19.77 & -22.31  & 38.45 & 0.2905 \\
L24 &  -21.28 & -23.63 &  39.04 & 0.3155 \\
S4  & -22.56 & -25.15 &  40.91 & 0.7125 \\
S5  & $-22.52^{[2]}$  & $-24.68^{[2]}$ &  40.71 & 0.3493 \\ 
S10 & -22.47 & -24.01 &  40.54  & 0.9923 \\ 
S11 & $-21.79^{[2}$ & $-23.70^{[2]}$ &  38.80 & 0.168 \\ 
\hline
\end{tabular}

[1] Not observed in $K^{\prime}$. $M_K$ is estimated by extrapolating the
model SED fitted to the observed $B-R$. [2] Munn et al. (1997) 
$B$ and $R$ mags (Table 1) used in estimate.
\end{table}

Concerning the radio luminosities, note that:

(i) The radio k-corrections (Appendix 1) 
are dependent on the assumed Condon (1992)
SED, in which the thermal ($\nu^{-0.1}$) component is 0.34 of the total flux at
10 GHz.  This fraction is consistent  with the mean  of $0.30\pm 0.05$
for spirals in the Shapley-Ames catalog (Niklas, Klein and Wielebinski
1997), but  the thermal fraction  is $>0.45$ in  15 per cent  of these
galaxies, and in the top 6 per cent reaches 0.60--0.75. If a galaxy at
$z=0.75$ has a thermal-component fraction  of 0.68 at 10 GHz, our
rest-frame 
$L_{8.44}$ would be an overestimate by 0.06 dex. 

(ii) Carilli (2000) predicts  a radio  luminosity per unit
SFR 2.35  times higher  than Condon  (1992), on the  basis of the
observed radio/FIR  flux ratio and FIR
emission  from  a model  starburst. 
 This estimate  was based  on the  extreme assumption  that the
fraction of the bolometric emission going into dust emission is unity,
which may be approximately  valid for dusty starbursts,
but in our models for  normal spirals this fraction is only 
$\sim  50$ per cent, which would reduce the radio flux per unit SFR 
almost to  the Condon (1992) value. We consider both models, as it seems
likely that real $L_{rad}$/SFR ratios would lie in the range between the
two, perhaps depending on the type of galaxy.

Figure 9 shows the radio/optical luminosity ratio, in the form of the
 restframe flux ratio,

\noindent  ${\rm log}~[F_{\nu}(8.44)/F_{\nu}(B)]=
{\rm log}~L_{8.44}- {\rm log}~(8.44\times 10^9)
 -[40.08-23.0+{\rm log}~3631-0.4M_B]$

\noindent $={\rm log}~L_{8.44}-30.566-0.4M_B$

against $M_B$. These ratios
are compared with (i) our Im model (approximately constant-SFR)
at $z=0$ (ii) the same model with the additional starburst component,
 increasing the total SFR by a factor 8.  
Figure 10 shows the ratio ${\rm log}~(F_{\nu}(8.44)/F_{\nu}(K^{\prime}))$ 
against $M_K$; this,  as the $K^{\prime}$-band
  will be less sensitive to 
starbursts and dust than $B$, 
should give a more direct indication of the SFR per unit stellar
mass.
We find that the redshift sample consists of   galaxies with
 a very wide, $\sim 2.5$
 order of magnitude, range  of radio/optical ratios, 
indicating a diverse mix of normal galaxies,
moderate and extreme starbursts, and AGN.

\begin{figure}
\psfig{file=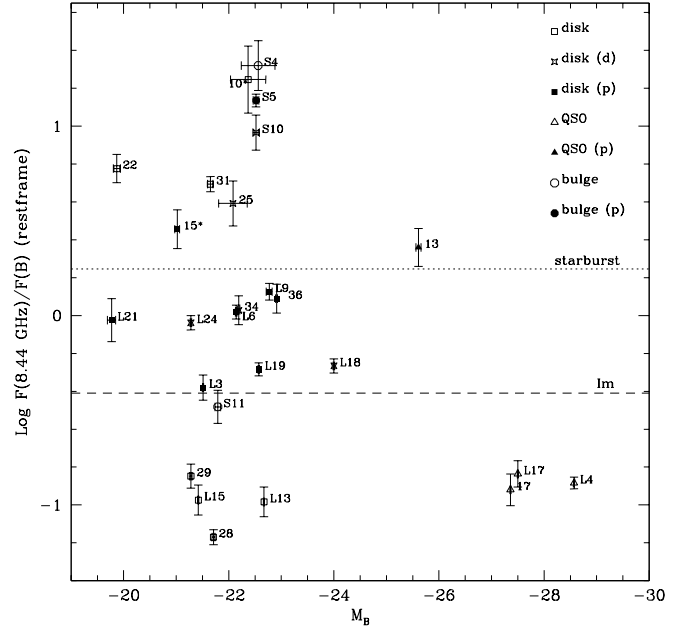,width=90mm}
\caption{Ratio  of rest-frame 8.44  GHz to  $B$-band flux  against $B$
absolute magnitude, for radio-detected galaxies. The dashed line shows
the flux ratio of the Im model  at $z=0$; the dotted line the Im model
with the addition of a starburst, increasing the SFR by factor 8.}
\end{figure}

\begin{figure}
\psfig{file=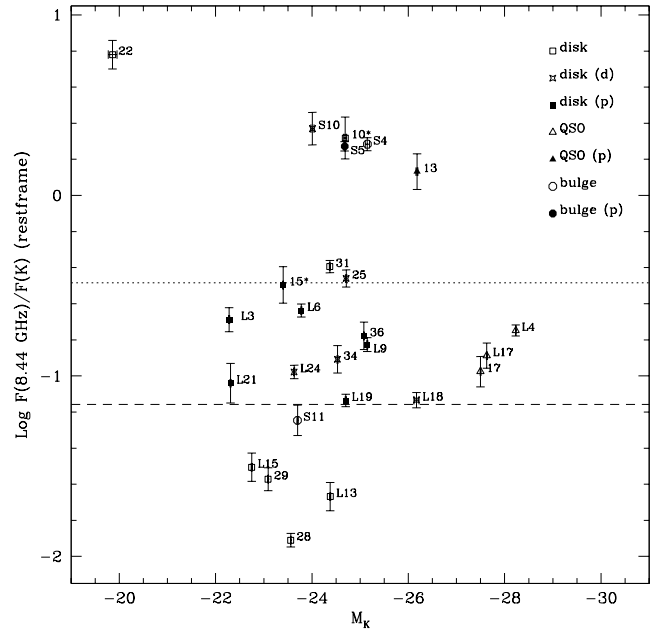,width=90mm}
\caption{Ratio of rest-frame 8.44 GHz to $K^{\prime}$-band flux against $K^{\prime}$ absolute magnitude. Models as in Figure 9.} 
\end{figure}
\section{Interacting Pairs}
We identify 10 close pairs within the redshift sample. 
  With our  chosen detection  criteria, 8  of these  pairs were
detected as separate galaxies on  both the LRIS and $K^{\prime}$ image
and the  radio source is assumed  to be the component  with the lowest
value of  $P$ (Section 4.1),  hereafter the `primary'. 
16V36 and L21  were detected as single sources on  the LRIS image but
are visibly double nucleus systems, and detected as pairs
on the  higher resolution NIRC images (the  $K^{\prime}$ magnitudes in
Table 1 are for the two components combined).

16V36 shows prominent tidal tails, confirming it is a merger.
The two  components of 16V15*  are verified by
spectroscopy  to be  at almost  identical redshifts,  as have  the two
bright  galaxies  16V28/9. In  the  case of  the  other  pairs a  firm
confirmation of interaction may require further (e.g.
WFPC2 or STIS) data. However, the statistical
significance of their close pairing can be estimated
 by calculating a pair probability $Pp$, again using the
formula $$Pp=1.0-{\rm  exp}(-\pi r^2 n(m))$$ where $r$  is the 
separation of  the two nuclei (as fitted  with SExtractor),
 $n(m)$ the  integral number  count (surface  density) of
galaxies  brighter than  the secondary  (fainter) galaxy  of  the pair
(magnitude $m$). $Pp$ is then  the probability that a companion galaxy
of similar or  greater brightness would, entirely by  chance, lie at a
similar or  closer distance  from the galaxy  identified as  the radio
source.

Table 4  gives the separations of  the galaxies, $r$, in  arcsec and in
$h_{50}^{-1}  \rm kpc$,  and the  $R$ flux  ratio (primary/secondary).
16V13 has three  companions detected as separate  objects on the
LRIS image, and parameters are given for each ('B', 'C' and `D').
\begin{table}
\caption{Close (likely  merging/interacting) pairs of  galaxies in the
sample of  radio detection  with redshifts: projected  separation $r$,
observed  flux ratio of  secondary to  primary (assumed  radio source)
galaxy  (in  the  $R$-band  unless  otherwise  stated)  and  estimated
probability $Pp$ of the pair occurring by chance.}
\begin{tabular}{lcccc}
\hline
Source & \multispan{2} Pair separation & $R$ flux & $Pp$ \\
\smallskip
       &  arcsec & $h_{50}^{-1}$ kpc & ratio & \\
16V13 (B) & 4.94 & 58.4 & 0.116 & 0.211 \\
16V13 (C) & 4.84 & 57.2 & 0.145 & 0.213 \\
16V13 (D)  & 4.13 & 48.8 & 0.086 & 0.223 \\
16V15*  & 2.50 & 21.1 & 1.117 & 0.004 \\
16V28/9 & 65.1 & 94.9 & 0.698 & 0.023 \\
16V36 & 1.84 & 14.0 & $0.912^{[1]}$ & 0.001  \\
L3 & 6.57 & 43.1 & 0.205 & 0.041 \\
L6 & 3.05 & 26.0 & 0.347 & 0.013 \\
L9 & 4.47 & 44.8 & 0.429 & 0.064 \\
L19 & 2.58 & 19.5 & 0.395 & 0.004 \\
L21 & 2.83 & 17.3 & $0.310^{[1]}$ & 0.030 \\
S5 & 6.06 & 42.0 & $0.052^{[1]}$ & 0.018 \\
\hline
\end{tabular}

[1] $K^{\prime}$ flux ratio
\end{table}
Excluding 16V13, the probability of each of these pairs being a chance
coincidence is  $<7$ per  cent.  In the  case of 16V13,  the companion
galaxies are faint and 
there is over a 20 per cent probability of each being 
a chance coincidence, but the likelihood of all three would be 
$\sim 1$ per cent. 

The mean
 luminosity  ratio of these pairs 
 is  $0.40\pm   0.11$,  and  the  mean  projected
 separation $41\pm 7$ $h_{50}^{-1}$ kpc, reducing to $36\pm 5$ if
 the  wide  16V28/9 pair  (which  are  not  obviously interacting)  is
 excluded. This mean separation is similar to that of both interacting
 FRII radio  galaxies (Roche  and Eales 2000b)  and of local
 radio-selected  starbursts (Smith,  Herter  and  Haynes
 1998), but greater (although marginally)  than the
 $19.1\pm 2.1$ $h_{50}$  kpc estimated for  merging ULIRGs 
(Clements  et  al. 1996).  
\section{Spectral Line Equivalent Widths}

\subsection{$\rm H\gamma$ and $\rm H\delta$}  
Delgado,  Leitherer  and Heckman  (1999)  model
Balmer   line   EWs   in   starburst  and   post-starburst   galaxies.
Star-forming  nebulae produce  $\rm  H\beta$, $\rm  H\gamma$ and  $\rm
H\delta$  emission  (with $\rm  H\beta$  the  strongest)  and stars 
(especially  spectral type  A) absorb these  lines,  so the
observed EWs are the difference ${\rm EW}_{em}-{\rm EW}_{abs}$. During
the burst the emission  dominates, but if star-formation is halted
it rapidly declines to zero (within $\sim 15$ Myr), and absorption
becomes dominant, peaking (${\rm EW}_{abs}\sim 13\rm \AA$) after $\sim 500$
Myr. Figure 11 shows a Delgado et al. (1999)
model for an instantaneous, solar-metallicity burst --  for $\rm H\beta$
we plot the emission and absorption components separately, and show
total EWs for
$\rm H\gamma$ and $\rm H\delta$. 
In a constant-SFR model (Delgado et al.  1999), $\rm
H\delta$  is  seen in  emission  for the  first  $\sim  100$ Myr. If 
steady star-formation continues  Gyr,  i.e. as in normal
spirals, it becomes an  absorption line, but remains weak compared to  that
in  post-starbursts. 

 Poggianti and Wu
(2000) define  a class  of `e(a)'  galaxies, with  both  strong Balmer
absorption, $\rm  EW (H\delta)>4 \AA$, and some  moderate emission in,
e.g., $\rm [OII]3727\AA$. Galaxies with e(a) spectra may be 
post-starburst     galaxies     with     some    residual,     ongoing
star-formation. However, Poggianti, Bresssan and  Franceschini (2000)
find   that  e(a)  spectra   can  also   be  produced   by  active
starburst galaxies if the most recently formed stars are
hidden by dust (`age-dependent extinction') (Section 10.3)

All non-QSO  galaxies in  our redshift sample  show $H\beta$
and/or $\rm [OII]3727\AA$ emission.  $H\gamma$ is seen in absorption
 for 16V31 and emission for 16V22 and 16V15*, but in other galaxies is
consistent with zero.   $\rm H\delta$ is seen in  absorption in 16V25,
16V31, L9,  L18 and  S4, but is  possibly a  weak
emission line  in 16V10*. These emission lines  indicate
 that 16V15*, 16V22 and 16V10*, at least,  are
actively starbursting. For  16V25 and S4, $  \rm EW(H\delta)<4\AA$,
which is consistent  with normal spirals. The  larger $\rm EW(H\delta)$
of 16V31 ($8.14\pm 1.10\AA$), L9 ($4.44\pm 1.61\rm \AA$)
and L18 ($5.22\pm 0.20\rm \AA$) classifies these as e(a) galaxies, and, if 
interpreted in terms of a simple  post-starburst model (Figure 12)
  corresponds to respective intervals  27 Myr, 7.6 and 8.7 Myr
 after the truncation of star-formation. 
\begin{figure}
\psfig{file=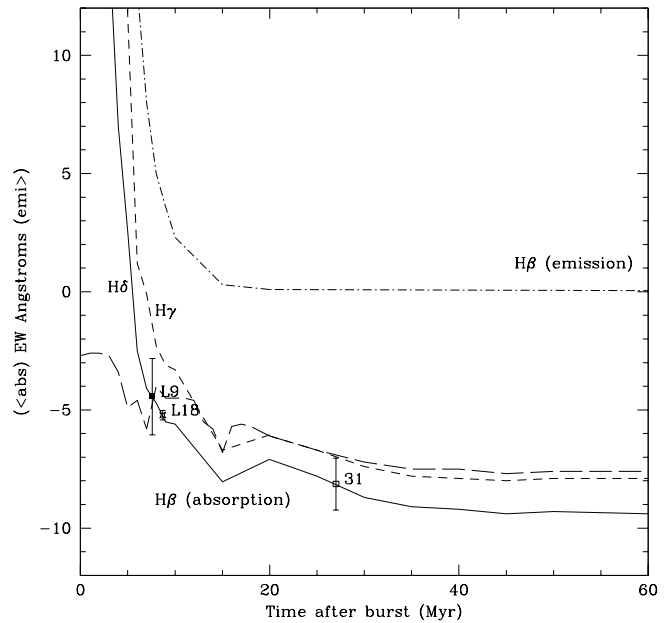,width=90mm}
\caption{Delgado et al. (1999)  model for the post-starburst evolution
of Balmer lines (emission  positive, absorption negative), showing the
total  (emission-absorption) EWs  for $\rm  H\gamma$  (short-dash) and
$\rm  H\delta$  (solid  line)   and  for  $\rm  H\beta$  the  emission
(dot-dash)  and  absorption (long-dash)  EWs  separately. The  plotted
points are the $\rm H\delta$ EWs  of the three `e(a)' galaxies, at the
starburst ages best fitting the model.}
\end{figure}
\subsection{$\rm H\beta$}
We observe $\rm H\beta$ emission for 9 galaxies.
The  luminosity, $L(\rm H\beta)$, is a good tracer of
the SFR on very short timescales - it takes only  $\sim 8$
 Myr to  fall by two
orders  of magnitude if star-formation ceases 
(Delgado  et al. 1999).  However, the observed  $\rm
H\beta)$ flux
is  affected  by  both  dust  extinction  and  the
superimposed absorption of the stars; the latter reduces the EW by 
 $\sim 2$--$5\rm \AA$ in normal spirals (Guzm\'{a}n et al.  1998).

 Kennicutt  (1983)  derived   a  relation  $L(\rm  H\alpha)=7.02\times
 10^{41}~SFR(>10M_{\odot})$ ergs $\rm s^{-1}$,  in the absence of dust
 $\rm H\alpha=2.86 H\beta$ (Guzm\'an et al. 1997), and for our assumed
 IMF the mass fraction of  $M>10M_{\odot}$ stars is 0.2105 (0.1267 for
 the Salpeter  IMF). Combining these gives  $L(\rm H\beta)= 5.17\times
 10^{40}~SFR$ ergs $\rm s^{-1}$ (where the SFR is for all masses).

To estimate $L(\rm H\beta)$ we first estimate an AB absolute magnitude
for the $4861\rm \AA$ continuum, $M_{4861}$, using the observed $B$ and
$R$ magnitudes (as in Appendix 1). A correction for stellar absorption
is added to 
all the  observed  $\rm  EW(H\beta)$, to give $\rm EW(H\beta)_{corr}$.
 The stellar absorption is assumed to be 
$2.7\rm \AA$  -- the mean for $z<0.7$ star-forming galaxies
in the Canada-France Redshift Survey  (Hammer et al. 1997) -- for all
except the e(a) galaxies, for which we estimate 
$6.90\rm \AA$ for 16V31 and $4.72\rm \AA$ for L9,  
from the Delgado
et al. (1999) post-starburst  models best fitting their $EW(\rm H\delta)$.
The line  luminosity is then

\noindent  $L({\rm H\beta})=10^{40.08}\times 10^{-19.44-0.4M_{4861}}
\times c/\lambda^2 \times {\rm EW(H\beta)_{corr}
}$ ergs $\rm s^{-1}$,

\noindent ${\rm log}~L(\rm H\beta)=20.64-0.4M_{4861}+ {\rm log}(1.2686\times 10^{11})+
{\rm log}~{\rm EW(H\beta)_{corr}}$

 (Table  5). Also given are the
$H\beta$    and   radio   (Condon    1992)   SFR    estimators,   $\rm
L(H\beta)/(5.17\times  10^{40})$  and $L_{8.44}/(3.66\times  10^{37})$
(both scale with $H_0$ as $h_{50}^{-2}$). The Carilli (2000) model
with our IMF gives a lower SFR of $L_{8.44}/(8.60\times 10^{37})$, which
can be regarded as a lower limit.  With  the standard  Salpeter IMF,  the SFR
estimators  increase  by factors  of  1.66  with  no change  in  their
ratio.  Combining these models (with no dust extinction) predicts ${\rm
log}~L(H\beta)={\rm log}~L_{8.44}+k$  where $k=2.78$ (Carilli 2000)
to $k=3.15$ (Condon 1992).
\begin{table}
\caption{Stellar-absorption  corrected  $\rm   H\beta$  EW,  log  $\rm
H\beta$ luminosity, and  SFR estimates from $\rm H\beta$  and 8.44 GHz
luminosities. SFR(8.44) 
is given  for the Condon  (1992) relation  and is
lower by a factor
of 2.35 for the Carilli (2000) model.} 
\begin{tabular}{lcccc}
\hline  Galaxy  &   $\rm  EW(H_{\beta})_{corr}$  &  ${\rm  log}~L({\rm
H\beta})$ & $\rm SFR(H\beta)$ & $\rm SFR(8.44)$ \\
\smallskip
      & $\rm \AA$ & ergs $\rm s^{-1}$ &   $M_{\odot}\rm yr^{-1}$ \\
16V15 & $12.29\pm 0.36$ & 41.30 & 3.87 & 74.1 \\
16V22 & $11.62\pm 0.34$ & 40.77 & 1.14 & 53.7 \\
16V31 & $9.99\pm 0.60$ & 41.53 & 6.57 & 229. \\
16V34 & $7.24\pm 0.24$ & 41.52 & 7.71 & 81.3 \\
16V36 & $7.52\pm 0.17$ & 41.86 & 14.1 & 182. \\
L6    & $10.53\pm 0.22$ & 41.68 & 9.22 & 75.9 \\
L9    & $13.33\pm 0.48$ & 42.12 & 25.3 & 174. \\
L21   & $15.00\pm 0.56$ & 41.02 & 2.05 & 7.76 \\
L24   & $6.75\pm 0.31$ & 41.14 & 2.68 & 30.2 \\
\hline
\end{tabular}
\end{table}
\begin{figure}
\psfig{file=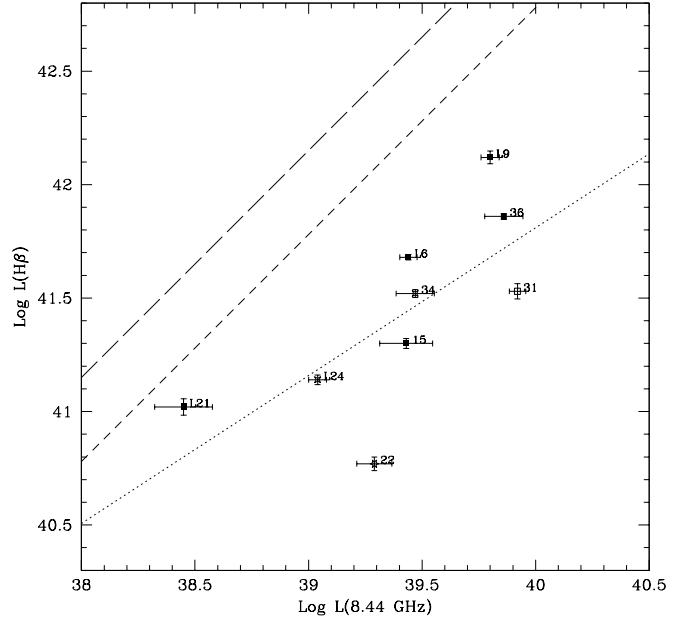,width=90mm}
\caption{Log $L(\rm H\beta)$  against $L_{8.44}$ in $h_{50}^{-1}$ ergs
$\rm   s^{-1}$   for  the   9   radio-detected   galaxies  with   $\rm
EW(H\beta)$. The short-dashed and long-dashed lines show the 
Carilli (2000) and Condon (1992) relations, with the assumption of no
dust extinction, and the dotted line the power-law
best-fitting the data.}
\end{figure}
Figure 11 shows $L(\rm H\beta)$ against $L_{8.44}$.
   The  two   luminosities   are  correlated   (correlation
coefficient  0.71),  with   a  best-fit  power-law  ${\rm  log}~L({\rm
H\beta})=15.73+0.652 {\rm  log}~L_{8.44}$, and a  scatter $\sigma({\rm
log}~L(\rm  H\beta))=0.30$. In  all  9   galaxies,  the  $L(\rm
H\beta)/L_{8.44}$ ratio is lower -- typically by an order of magnitude
-- than the models.
 At  the  mean ${\rm  log}
~L_{8.44}$    of     39.41,    $\rm    H\beta$     is
$2.5(39.41+k-41.43)=1.9$--$2.8\pm 0.25$ mag fainter than predicted.
We discuss this further in Section 10.5.

\subsection{[OII] $3727\rm \AA$}
The [OII] $3727\rm \AA$ emission line is also an indicator of the 
current SFR -- 
  Kennicutt (1992)       estimates       $L(\rm[OII])=3.15\times
10^{41}~SFR(>10M_{\odot})$ ergs $\rm s^{-1}$ (with no dust). For
the IMF adopted here, this becomes $L(\rm[OII])=6.63\times 10^{40}~SFR$
ergs $\rm  s^{-1}$.

We detected [OII] emission from 6 of the galaxies. Estimating continuum
absolute magnitudes  at $3727\rm \AA$, $M_{3727}$, by  the same method
as used above for $M_{4861}$, the luminosity

$L(\rm[OII])=10^{40.08}\times 10^{-19.44-0.4M_{3727}} \times c/\lambda^2 \times {\rm EW([OII])}$ ergs $\rm s^{-1}$,

${\rm log}~L(\rm[OII])=20.64-0.4M_{3727}+ {\rm log}(2.158\times 10^{11})+
{\rm log}~{\rm EW([OII])}$

is given in Table 6, with the SFR as estimated from  $\rm L([OII])$ and $L_{8.44}$. 
\begin{table}
\caption{$\rm [OII]3727  \AA$ EW (in $\rm \AA$),  log [OII] luminosity
(ergs  $\rm s^{-1}$),  and SFR  estimates (in  $M_{\odot}\rm yr^{-1}$)
from [OII] and 8.44 GHz luminosities, for galaxies with detected [OII]
emission.  SFR(8.44)  is given for  the Condon (1992) relation  and is
lower by a factor
of 2.35 for the Carilli (2000) model.}
\begin{tabular}{lcccc}
\hline
\smallskip
Galaxy & $\rm EW([OII])$ & ${\rm log}~L(\rm[OII])$ &
$\rm SFR([OII])$ & $\rm SFR(8.44)$ \\
16V10* & $7.37\pm 1.10$ & 41.43 & 4.06 & 1585. \\
16V25 & $8.78\pm 0.56$ & 41.38 & 3.67 & 269. \\
L9    & $15.45\pm 0.98$ & 41.90 & 11.9 & 174. \\
L18   & $5.68\pm 0.15$ & 41.99 & 14.7 & 219. \\
S4    & $7.54\pm 0.53$ & 41.55 & 5.32 & 2239. \\
S10   & $52.50\pm 0.84$ & 42.47 & 44.6 & 955. \\
\hline
\end{tabular}
\end{table}
S10 has a high $\rm EW([OII])$, typical of
 a late-type starburst, while the 
the other 5 have moderate EWs in the range of normal
Sb--Sc spirals (Kennicutt 1992; Sodre and Stasinska 1999).
\begin{figure}
\psfig{file=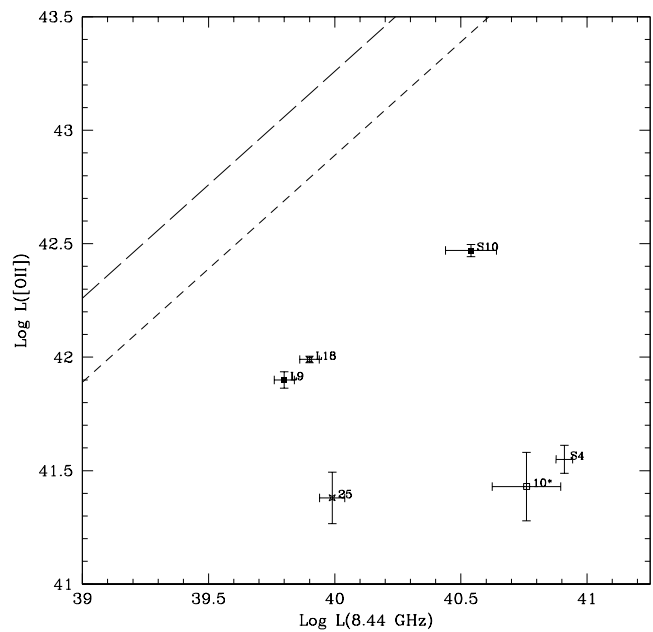,width=90mm}
\caption{Log $L({\rm [OII]3727\AA})$ against $L_{8.44}$, in
$h_{50}^{-1}$ ergs $\rm s^{-1}$, for the 6 radio-detected galaxies
with $\rm EW([OII])$. 
The short-dashed and long-dashed lines show the 
Carilli (2000) and Condon (1992) relations, with the assumption of no
dust extinction.}
\end{figure}
Figure 12  shows $L(\rm [OII])$ against $L_{8.44}$,  together with the
no-dust   models,  ${\rm   log}~L([OII])={\rm   log}~L_{8.44}+k$  with
$k=2.89$ (Carilli 2000) and $k=3.26$ (Condon 1992). With so few points
the luminosities are not  significantly correlated, but the mean ${\rm
log}~L($\rm [OII]$)-{\rm log}~L_{8.44}$ is  1.47 (with scatter
0.68), corresponding to 
$2.5(k-1.47)=3.6$--$4.5\pm 0.7$ mag extinction at $3727\rm\AA$. If
16V10* and S4 are excluded (on the grounds that they may have obscured
AGN) the mean log ratio is 1.90 (scatter 0.34), indicating
2.49--$3.4\pm 0.7$ mag.

\subsection{[OIII] $5007\rm \AA$}
The ratio of $\rm [OIII]5007\AA$  
to  $\rm H\beta$ flux (`excitation') is  a useful diagnostic
of the type of galaxy. In local galaxies, it typically increases from $\sim
0.2$  to 1--2  from Sb  to Im  types (Sodre  and Stasinska  1999), and
further to a
mean  of   $3.1\pm  0.3$  for   dwarf  HII  galaxies   (Guzm\'{a}n  et
al. 1997). The excitation of starburst galaxies 
may cover a wide range from
$\sim 0.3$ to $\sim 5.0$ (Tresse et al. 1996), and is
strongly dependent on metallicity; models predict it to increase
slightly from $0.1 Z_{\odot}$ to a peak at $0.25Z_{\odot}$, then fall by
factors 2--3  to $1Z_{\odot}$ and by  a further order  of magnitude to
$2Z_{\odot}$.   Higher excitations  of  $\sim 5$--15  are observed  in
Seyferts and are usually interpreted as evidence for an AGN.

[OIII] emission was observed for  the same 9 galaxies as $\rm H\beta$.
As described in Appendix 1,
 we estimate from the colours of each, the ratio of the continuum
$F_{\nu}$  at $\rm  5007\AA$ to  that  at $\rm  4861\AA$,
typically  $\sim  1.1$.  This ratio is multiplied by $\rm
EW([OIII])/ EW(H\beta)_{corr}$  (Table  5), giving  the
excitation (Table 7).

\begin{table}
\caption{ $\rm EW([OIII]5007)$ (in $\rm \AA$) and the ratio of [OIII] 
to absorption-corrected $H\beta$ flux (excitation), for
 galaxies observed at $\rm O[III]$.} 
\begin{tabular}{lcccc}
\hline
\smallskip
Galaxy & $\rm EW([OIII])$ & $F([{\rm OIII}])/F({\rm H\beta})$ \\
16V15 & $6.16\pm0.37$ & $0.576\pm 0.051$ \\
16V22 & $30.19\pm 0.54$ & $2.925\pm 0.138$ \\
16V31 & $6.29\pm 0.44$ & $0.683\pm 0.091$ \\
16V34 & $3.37\pm 0.27$ & $0.536\pm 0.061$ \\
16V36 & $2.28\pm 0.08$ & $0.325\pm 0.019$ \\
L6    & $3.81\pm 0.21$ & $0.418\pm 0.032$ \\
L9    & $6.74\pm 0.54$ & $0.542\pm 0.064$ \\
L21   & $40.19\pm 0.66$ & $2.744\pm 0.148$ \\
L24   & $2.70\pm 0.31$ & $0.461\pm 0.074$ \\
\hline
\end{tabular}
\end{table}
These excitations are bimodally distributed, with 7/9 galaxies 
forming
a group with
a mean of 0.51 and a small scatter
of 0.12, while 16V22 and L21 both have high excitations of 
$\sim 2.8$. None  have excitations  sufficiently  high  to  require an  AGN
contribution. The excitation is negatively correlated with $B$
luminosity (correlation coefficient 0.88). 
 16V22 and L21 are optically the least luminous galaxies in the
redshift sample, with $M_B>-20$. Their $M_B$, $L(\rm H\beta)$ and
excitations are similar to those of the $z<0.7$,
 low metallicity($\sim 0.4Z_{\odot}$) HII galaxies studied
by Guzm\'{a}n et al.  (1997). The excitations of the other 7 are in the lower
half of the range of more luminous ($L\sim L^*$), metal-rich
($Z_{mean}\sim 0.8$--$1Z_{\odot}$)
starburst and interacting spirals (Gallego et al.  1997; Pastoriza and
 Donzelli 2000).
 
Kobulnicky and Koo (2000)  describe an estimator of metallicity (O/H),
originally from  McGaugh (1991), based  on the ratio of  emission line
fluxes        $\rm        R_{23}=({[OII]3727+[OIII]4959+[OIII]5007\over
H\beta})$.  Assuming $\rm [OIII]4959=[OII]5007/2.9$  and the  [OII] to
[OIII] relation fitted  to HII galaxies by McCall,  Rybski and Shields
(1985), the $\sim 2.8$ excitation of 1622 and L21 corresponds to ${\rm
log}~R_{23}\simeq 0.825$ and  ${\rm log}(Z/Z_{\odot})\simeq -0.25$ and
the   $0.51\pm  0.12$  of   the  more   luminous  galaxies   to  ${\rm
log}~R_{23}\simeq  0.45\pm  0.05$  and  ${\rm  log}(Z/Z_{\odot})\simeq
+0.05\pm 0.03$.  
 The excitation/metallicity vs.  luminosity trend of these 9
galaxies ($z_{mean}=0.45$) appears
consistent with that seen for the $z\sim 0$--0.4
emission-line galaxies of Kobulnicky and Koo (2000).

\section{Kinematics from Emission Lines}
The  2D profiles of  spectral  emission lines may provide
information on the galaxy kinematics -- internal velocity dispersion
and rotation velocity. These can distinguish
between giant spirals (high dispersion and rotation), normal giant
ellipticals  (high  dispersion,  low  or  zero  rotation),  and  dwarf
galaxies (low dispersion, low or zero rotation), and furthermore
the rotation velocity of spirals is well correlated with mass and optical
luminosity. 
\subsection{Velocity Dispersion}
In many of our spectra  the emission lines are noticeably broader than
the instrumental resolution. Guzm\'{a}n et al. (1997), observing faint
compact galaxies  with the same spectrograph,  estimated  internal  velocity
dispersions ($\sigma_{i}$) as
$$\sigma_{i}=\surd({\rm     FWHM}^2-(3.1{\rm     \AA})^2)\times{c\over
2.35\lambda_0(1+z)}$$  where  $\sigma_i$  is the  rest-frame  internal
velocity dispersion of the galaxy, the line FWHM 
is from a  Gaussian fit to the strongest  emission line in the
spectrum,  $\lambda_0$ is the restframe wavelength  of this line,
and $3.1\rm \AA$ is the effective instrumental resolution of LRIS with
the 600 $\rm mm^{-1}$ grating. 

Applying this method to our spectra, using in most cases the $H_{\beta}$
line, we find a significantly positive 
$\sigma_{i}$ for all 14 non-QSO galaxies (Table 8). 
\begin{table}
\caption{Internal  velocity  dispersion $\sigma_i$  and  the shift  in
velocity  across  the galaxy  $\Delta  v$,  as  estimated from  the  
profiles of the strongest emission lines.}
\begin{tabular}{lcc}
\hline
\smallskip
Galaxy & $\sigma_i$ (km $\rm s^{-1}$) & $\Delta v$  (km $\rm s^{-1}$) \\
16V10 & $177\pm 28$ & $202\pm 67$ \\
16V15 & $90\pm 5$ & $40\pm 17$ \\
16V22 & $58\pm 2$ & $15\pm 7$ \\
16V25 & $151\pm 11$ & $164\pm 36$ \\
16V31 & $260\pm 23$ & $333\pm 81$ \\
16V34 & $225\pm 16$ & $450\pm 37$ \\
16V36 & $166\pm 6$ & $71\pm 38$ \\
L6 & $88\pm 3$ & $149\pm 28$ \\
L9 & $199\pm 12$ & $344\pm 22$ \\ 
L18 & $163\pm 5$ & $198\pm 17$ \\
L21 & $47\pm 2$ & $21\pm 9$ \\
L24 & $87\pm 14$ & $7\pm 26$ \\
S4 & $211\pm 18$ & $311\pm 59$ \\
S10 & $160\pm 3$ & $216\pm 12$ \\
\hline
\end{tabular}
\end{table}
\begin{figure}
\psfig{file=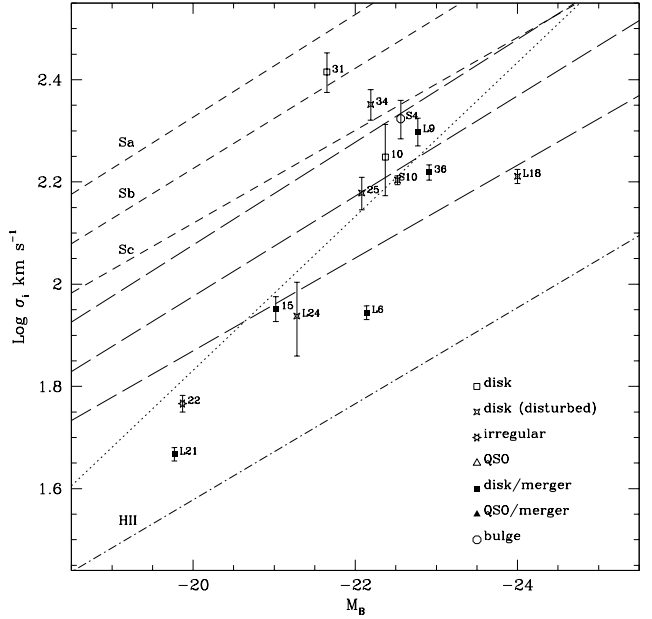,width=90mm}
\caption{Internal velocity dispersion ($\sigma_i$),
 as estimated from emission-line FWHM, 
against blue  absolute magnitude  $M_B$. The
dotted line is the power-law best fitting these data. The short-dashed
lines show the mean relations of rotation velocity (not $\sigma_i$) to
$M_B$  for Sa,  Sb  and Sc  spirals,  from Rubin  et  al. (1985).  The
long-dashed lines  show these same relations shifted  downwards by 0.25
dex to more directly compare with the observed $\sigma_i$
(see  text). The  lowest, dot-dash  line is  the Telles  and Terlevich
(1993) $\sigma_i$--$M_B$ relation for HII galaxies.}
\end{figure}
 
We find $\sigma_i$ to be correlated (correlation coefficient 0.75)
with  luminosity  (Figure 13);  the  best-fitting  power-law is  ${\rm
log}~\sigma_i=-0.1506M_B-1.181$,   with   a   scatter  $\sigma   ({\rm
log}~\sigma_i)=0.15$.  If these are disk galaxies,
 $\sigma_i$ will be related to the
rotation velocity  (on the flat  outer part of  the rotation
curve) $v_{rot}$, although  with a large scatter due  to the random
orientations. 

Using simulations Rix
et al. (1997) predict most  spirals in a randomly oriented sample will
have      $\sigma_i/v_{rot}=0.45$--0.9,      with      an      average
$\sigma_i/v_{rot}\simeq 0.65$.
 Rubin et al. (1985) give observed $v_{rot}$--$M_B$ relations 
for Sa, Sb and Sc
spirals (plotted on Figure 13).  If our galaxies are like 
local spirals, they would,
due to their random orientations, lie
an average of $\sim 0.19$ dex below these relations. 
 In addition,  galaxy evolution is reported to  shift the
local $v_{rot}$--$M_B$ relation  by up to  $\Delta(M_B)\sim -0.6$ mag  at these
redshifts (Vogt  et al. 1996), equivalent  to 
$\Delta(v_{rot})=-0.06$ dex at a given $M_B$.  Hence,  for  a  direct
comparison with our $\sigma_i$,  we also plot the Rubin
et al.  (1985) relations shifted down by $\Delta(v_{rot})=-0.25$
dex. Also shown is the  
Telles and Terlevich (1993)
$\sigma_i$--$M_B$  for HII  galaxies, which  have much  lower internal
velocities than spirals of the
same $M_B$ (see also Koo et al. 1995). 

 We find that the more luminous
( $M_B<-21.5$) galaxies generally have   $\sigma_i$    
consistent with the relations for  spirals, but the 
   low-luminosity 16V22 and L21    have  $\sigma_i$ below these
relations      
       and may be       more consistent with HII galaxies. 

\subsection{Rotation
 Velocity} A  second and more  direct measure of galaxy  kinematics is
 the shift  in an emission  line's centroid wavelength across the
 galaxy in the  spatial direction. The
 wavelength shift  $\Delta\lambda$ corresponds to  a rest-frame radial
 velocity,  $\Delta  v={c\Delta\lambda\over  \lambda_0(1+z)}$.  An
edge-on spiral with long axis along the spectroscopic  slit will
 approach
  $\Delta v\sim 2v_{rot}$,  but a sample of randomly oriented spirals
 will be closer to $\Delta v\sim v_{rot}$, with a large scatter.

On our 2D (spectral $x$ $\times$ spatial $y$) spectra, the majority of
the  galaxies show  diagonally slanted  emission lines.
  From each 2D spectrum,  we  extract a
series of  5--8 1D  spectra from narrow  strips of width  $\Delta y=4$
pixels, shifting the  strip aperture in steps of  $\Delta y= 2$ pixels
across the visible  width of the spectrum, and then  for each of these
narrow-strip  spectra  measure  (with  a Gaussian  fit)  the  centroid
wavelength  of  the strongest  emission  line. Figure 15
shows, for four of  the more 
strongly rotating galaxies,  the trends in  radial velocity with $y$, 
which resemble  the rotation curves of known spirals
(e.g. Rubin et al. 1985).
\begin{figure}
\psfig{file=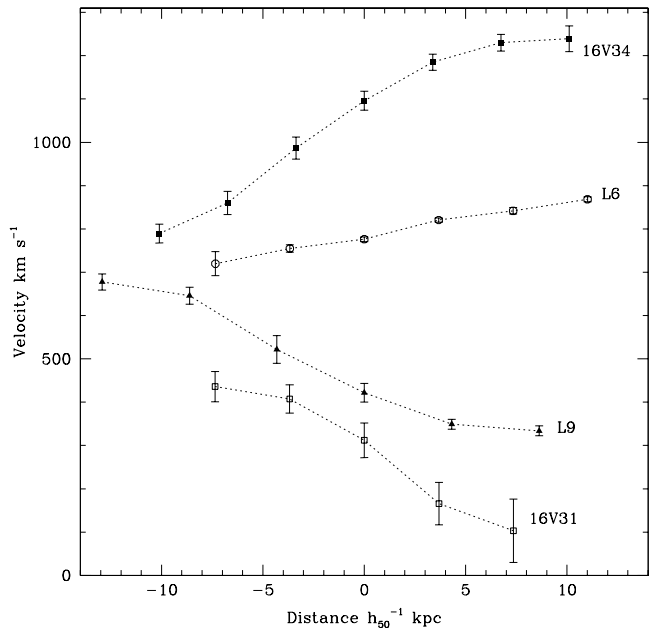,width=90mm}
\caption{Rest-frame radial velocity (arbitrarily offset on the $y$ axis) as measured from the centroid of the strongest emission line, as a function of distance across the galaxy (in $h_{50}^{-1}$ kpc), for
 four rotating galaxies.}
\end{figure}

 For each  galaxy, we  measure the difference  in maximum  and minimum
 line  $\lambda$ across the spectrum, and in Table 8 
give the corresponding radial
 velocity difference, $\Delta v$. This
 shows some correlation with luminosity
 (correlation  coefficient  0.69), with  a  best-fit  power law  ${\rm
 log}~\Delta  v=-0.3429M_B-5.501$ (Figure 16). The 
power-law  is steeper  than for
 $\sigma_i$, but the two velocities are about equal for the more
 luminous 
galaxies
 ($M_B\simeq -22.5$). Much of the scatter $\sigma ({\rm
 log}~\sigma_i)=0.41$ is undoubtedly due
to the random orientations  -- 16V34 and 16V31
appear almost optimally aligned while e.g. 16V15 and 16V25
are almost face-on,  and in 16V36 the slit passes  through only one of
the two nuclei. 

 Again  the $v_{rot}$--$M_B$  relations  of Rubin  et  al. (1985)  are
plotted.  About 8  or 9 of the 14 galaxies  are consistent with these,
assuming $\Delta v\sim v_{rot}$.  The steepness of  our 
fitted  correlation, compared to the $\sim 0.1$ slope 
of these single-type  relations, is most likely due to  the lower luminosity
galaxies in our sample being later Hubble types and/or HII galaxies.
\begin{figure}
\psfig{file=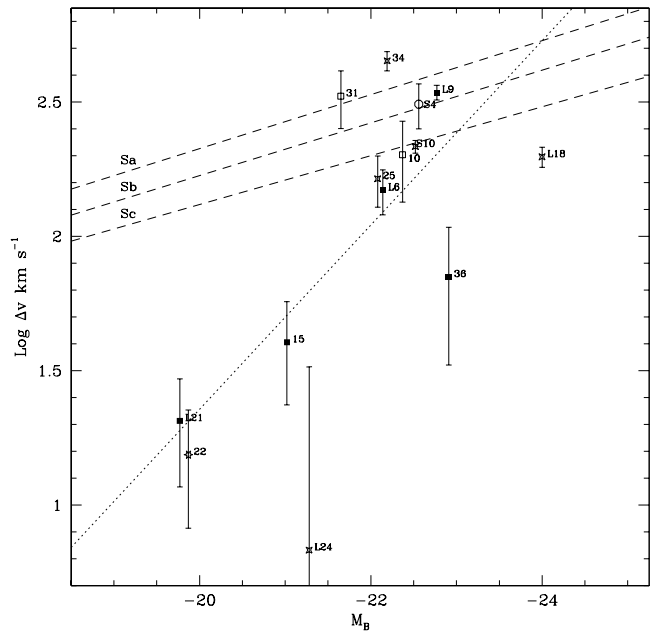,width=90mm}
\caption{The  radial velocity  shift across  each galaxy,  $\Delta v$,
plotted against blue absolute magnitude  $M_B$. The dotted line is the
best-fit  power-law.  The  dashed  lines show  the  mean relations  of
rotation velocity  to $M_B$ for Sa,  Sb and Sc spirals,  from Rubin et
al. (1985)}
\end{figure}

 We conclude that 12/14 galaxies are consistent with spiral
kinematics, while 
the low-luminosity 16V22 and L24 are low in both $\sigma_i$ and
$\Delta v$ and appear more consistent with HII galaxies (Guzm\'{a}n et
al. 1997).  None of these 14 show the kinematics of normal 
giant ellipticals,
which  have high  velocity dispersions,  $\sigma_i\simeq 250$  km $\rm
s^{-1}$,      but       much      lower      rotation      velocities,
$0<v_{rot}<0.4\sigma_{i}$.  S4,  although  bulge-profile, gave  a
high $\Delta v$ which might be evidence of an 
ongoing merger.

\section{Discussion -- the Nature of $\sim 10\mu\rm Jy$ Sources}

\subsection{Comparison with Hammer et al. (1995)}
 Hammer  et.
al.  (1995) investigated  a sample of galaxies
to  slightly shallower 
but comparable limits of 
$F(5.0~\rm GHz)\simeq  16\mu Jy$ and $I_{AB}=22.5$. Firm spectroscopic
IDs included  one QSO, a possible radio galaxy, 
 5 ellipticals, 6 spirals and 5
 emission-line galaxies. Their results agree with ours in
 that the most prominent  type of galaxy were  $L\geq L^*$ 
spirals, with 
  moderate [OII] emission and often Balmer absorption lines. 
 Hammer et al. (1995) report a  higher proportion of ellipticals
 than our 3/27, but 3 of their 5 spectroscopically
 confirmed ellipticals are at $z=0.99$, suggesting  the
elliptical fraction is enhanced by a cluster.
There is evidence
 that 
$z\geq 1$ ellipticals are even more clustered than the present-day E/S0
 population (Daddi et al. 2000), so large 
 field-to-field variations in their numbers 
 would be expected. Furthermore, at the less deep radio flux limit of
Hammer et al., radioluminous
ellipticals (AGN) are likely to be  more numerous relative to
star-forming galaxies.

Of the 5 galaxies described by Hammer et al. as `emission-line', 4 are
sub-$L^*$ galaxies with line ratios more consistent with 
Seyferts or LINERs than pure starbursts.
 In contrast, we concluded that our 
two sub$-L^*$ galaxies  
had line ratios  within the range of HII galaxies. However,
these variations are not surprising considering that line-ratio diagnostic
plots show a  great
concentration of low-luminosity galaxies around the HII/Seyfert/LINER
dividing node (Tresse et al. 1996). The fifth is a 
 high redshift ($z=1.16$) galaxy, with strong [OII] emission and no
direct evidence for an AGN, and   might be similar to
our S10.

\medskip

\subsection{`Normal' Galaxies}
Some $\sim 20$ per cent of our sample consists of apparently `normal' galaxies at
lower redshifts. The disk galaxies 16V28, 16V29,
L13 and L15 and the elliptical S11, are optically
bright ($R<19$),  low redshift ($z_{mean}=0.15\pm  0.05$), $L\sim L^*$
luminosity  galaxies with  relatively low  radio/optical  flux ratios,
consistent with non-starburst spirals or normal E/S0s. None appear visibly
disturbed or interacting. 

\subsection{QSOs}
Some $\sim 15$ per cent of the sources
are  identified from their spectra  as QSOs. These sources,
16V13, 16V17, L5 and L17,
 are all high
redshift  ($1.6<z<2.6$),  optically  very luminous  ($M_B\sim  -25.5$)
objects with starburst-like $B-R$  colours.
16V17, L4 and L17  appear stellar and have
relatively  moderate radio/optical flux  ratios, intermediate
between the normal and starburst galaxies. 

The QSO 16V13  differs from the others in that
(i) its radio/$K$  band luminosity
ratio  is a  factor  $\sim 8$  higher  (Figure 10),  (ii) optical  and
$K^{\prime}$ images  show the compact  bright source to  be associated
with a  lower surface brightness  `tail' and three  blue 
($B-R\simeq  0.3$) companion galaxies;
(iii)  the spectrum  shows a  doublet  of very  strong ($\rm  EW\simeq
16$--$19\rm \AA$) narrow MgII  absorption lines, and (iv) Windhorst et
al.  (1993) found  the radio  source to  be extended  --  elongated by
$12.6\pm 3.0$ arcsec  on a position angle $133^{\circ}\pm  25$. It may
therefore be a more complex object, e.g. a radio QSO with
jet-triggered star-formation. If the radio emission is primarily
from starbursting, the Dunne, Clements and Eales (2000) relation 
of $F(850\rm \mu m)/F(/\rm 1.4~GHz)$ to $z$ for star-forming galaxies predicts
$F(850\rm\mu m)\simeq 17 mJy$, which would be easily detectable with SCUBA. 
\subsection{Obscured AGN (probable)} 
Two of the three ellipticals, S4 and S5,
 have a very high  $L_{8.44}$, which  if due  to  star-formation, would
require
$\rm SFR\geq  1000M_{\odot} yr^{-1}$. These galaxies appear  to be
regular giant  ellipticals without visible point-sources  or knots.
It seems more  likely that their  radio emission is produced  by obscured
AGN. Radio bursts from AGN may be triggered by even minor interactions. 
S5 appears to be the central galaxy of a group or cluster
 (Weistrop et al. 1987), with a much smaller dwarf
galaxy  $\sim  40$ kpc  from  the  nucleus.  S4 is  not  visibly
interacting on these images, but its spectrum showed a high velocity
gradient, which may be evidence for an interaction. 

It is  also possible  that 16V10* contains  a radio  AGN, as it  has a
similarly  high $L_{8.44}$  and the  radio  source is  reported to  be
elongated (Windhorst et al. 1993).

\subsection{Starburst Galaxies}
More than half of our sample consists of  disk
galaxies   with  high  radio   luminosities, suggesting   powerful
($\sim \rm 100 M_{\odot}yr^{-1}$) starbursts; 16V15*, 22, 25, 31, 34, 36, L3, L6, L18, L19,
L21, L24, S10 and possibly 16V10*. 
The radio/$K$  luminosity ratio -- 
an approximate  indicator of the  ongoing/recent SFR  relative to
the total  stellar mass -- varies greatly ($\sim  2.5$ dex)
within
 the sample as a whole, is not
obviously  correlated with  optical luminosity,  and  its distribution
extends $\sim  1.5$ dex above  a constant-SFR model (Figure  10). This
appears   to  reflect   the  underlying   distribution   of  starburst
intensities, as the optically-based Guzm\'{a}n et al.  (1997) plot of
SFR/mass against galaxy mass is very similar.

{\bf Interactions:}  Interaction-triggered starbursts appear  to be the
primary cause of enhanced radio emission from disk galaxies. 
We identify 11 of our redshift sample as being
in close pairs, and several others appear disturbed. 
For  one   pair,  16V15*/16V15B, we  have
spectra for both components, confirming the redshifts are the same; 
another galaxy, 16V36, has prominent tidal tails and a double nucleus,
leaving  no  doubt that  it  is a  merger.   S10,  with the  strongest
emission lines, is very asymmetric -- the
nucleus is off-centre  by $\sim 1/4$ of  the semi-major axis --
almost certainly from an interaction.

Serjeant et al.  (2000) imaged four radio-selected
starburst galaxies at $z\sim 0.2$ with WFPC2, and found all four to be
interacting  or  disturbed, and  two  to  be  very asymmetric.  On
this basis it is
likely that higher  resolution imaging would reveal evidence
of interaction in other galaxies in our sample.

The mean projected
 separation of the pairs is a relatively wide
$36\pm5 h_{50}^{-1}$ kpc, which 
implies that, in many mergers,  the SFR and radio emission
 peak  well  before nuclear
 coalescence (36  kpc is 117 Myr  at 300 km $\rm  s^{-1}$).
The wide range
 of separations
 is consistent with merger simulations where the maximum SFR may occur 
at nuclear
coalescence or several $\times 10^8$ yr earlier, depending on the
structure of the galaxies (Mihos and Hernquist 1996) and their 
relative orbital configurations (Bekki and Shioya 2000).

{\bf Excitation:}  We estimated the  $\rm [OIII]5007$ to  $\rm H\beta$
flux ratio (excitation) for 9 of the galaxies, and found seven to 
have moderate excitations, averaging 0.51,
implying the  high ($Z\sim  1.1 Z_{\odot}$) metallicities  of $L\geq
L^*$ spirals. Two with lower optical luminosities ($M_B>-20$)
have much
higher excitations of $\sim 2.8$, consistent with low metallicity ($Z\sim
0.5  Z_{\odot}$) HII  galaxies.  This diversity  of  spectral types  is
similar to that in the Hammer et al. (1995) sample, although
 none of these 9 galaxies have the very 
high  excitations  ($\geq  5$)   indicative  of  AGN.   The  
metallicity -- luminosity relation appears consistent with that of 
optically selected
galaxies and hence with the radio sources being formed from `normal'
spirals and irregulars, e.g. through interactions. 

{\bf Kinematics:} Using the strongest emission lines in our spectra we
estimated internal velocity dispersions and rotation velocities for 14
galaxies.  Most of the sample were 
consistent with  the kinematics ($v_{rot}\simeq 100$-400  km $\rm s^{-1}$)
of massive spirals. The
lowest  optical luminosity galaxies,  16V22 
and L21 had much lower velocity dispersions and rotation velocities,
$<60$  km  $\rm  s^{-1}$,   consistent  with  low-mass  HII  starburst
galaxies.

\subsection {Radioluminous Starbursts: Dust and Age Effects}
The  SFRs estimated from radio
 luminosities are consistently about an order of magnitude higher than those
 indicated by the emission line ($H_{\beta}$ and/or $\rm [OII]3727\AA$)
 luminosities. The most obvious explanation is dust
 extinction of the emission lines.
The typical shortfall in $H_{\beta}$ flux, interpreted in this
 way, implied
$A(H\beta)=1.9$--2.8 mag, with similar results for  $\rm [OII]3727\AA$.
This is a common finding in radio-luminous starbursts, e.g.
 Smith et
   al.  (1996) found  the ratio  of $\rm  Br\gamma$  ($2.166\mu\rm m$)
   to thermal radio emission to be only 4--18 per cent of that
exected without dust, in starburst galaxies (mostly interacting) of
   similar radio luminosity to our sample. Furthermore, in the 
10  Wolf-Rayet galaxies of  Beck, Turner  and Kovo  (2000), and  the 4
 starburst galaxies of Serjeant et al. (2000), the
   $H\alpha$ luminosities  correspond to
   SFRs a factor $\sim 10$ lower than the radio luminosities.

Despite this, our radio IDs are not in general
significantly redder than optically-selected galaxies (Figure 5), and some
have very blue, starburst colours.  This discrepancy 
can be explained if 
the dust is inhomogeneously distributed and
produces an `age-dependent' extinction  -- several  magnitudes 
for the most newly formed stars ($<10^7$ yr), progressively decreasing
for  stellar  populations of  increasing  age
(e.g. Poggianti and
Wu 2000; Poggianti  et al. 2000; also the merger model of Bekki and
Shioya 2001). $H\beta$  emission,
tracing star-formation over the previous few Myr, would be  greatly
attenuated, whereas slightly older stars emerging from high-extinction regions
can produce blue colours and, eventually, 
$H\delta$ absorption.  Hence galaxies with ongoing, reddened
starbursts could still appear very blue and may have
e(a) spectra.

However, an estimate of dust extinction from $H_{\beta}/L_{8.44}$
 would be strictly valid only if the SFR  is relatively  constant over
   $>10^8$ yr, as a consequence of 
$L(H_{\beta)}$ and $L_{8.44}$ tracing star-formation over different
 timescales, $<7$ Myr and $\sim 70$ Myr.  Shorter  starbursts would
 cause their  ratio to fluctuate. 
  Glazebrook et al. (1999)
   and Sullivan et al.  (2000) find star-forming galaxies to show
   a very wide scatter in the ratio of $H_{\alpha}$ to
   UV continuum flux, and 
  attribute this to most of the 
star-formation  occurring  in  repeated  $\sim  50$  Myr   bursts. 
Thornley et al.  (2001) report that variations  in the 
 mid-infra-red line ratios of starburst galaxies
 imply  that the most massive stars are formed in 
even  shorter ($\sim  10$  Myr)  bursts.

Sullivan  et al.  (2000)  estimated  SFRs from  the  UV continuum  and
 $H\alpha$ luminosity of UV-selected  galaxies. Even after taking into
 account a greater extinction for the  star-forming nebulae ($A_N$) 
compared to the whole stellar population
 ($A_s$),   assuming  $A_s=0.44   A_N$  
(see   Calzetti   1997),  they
 consistently found the $H\alpha$-estimated  SFRs
 to be lower than the UV  estimates. 
 In contrast, Glazebrook
 et  al.   (1999)  had found  the opposite result, higher  $H\alpha$
 SFRs,  for
 optically-selected  galaxies. The discrepancy  was attributed  to the
 wide intrinsic scatter in $H_{\alpha}/L_{UV}$ combined
 with selection effects. In the same way, 
a radio-selected sample may be 
   biased towards a lower than `normal' (i.e. steady-SFR)
mean $H_{\beta}/L_{8.44}$, by the inclusion of 
 many  recent
 post-starbursts.

In our spectroscopic sample we identify 3 `e(a)' galaxies,  L9, L18 and
16V31, out of 7  where the observed $\lambda$ range would include 
$\rm H\delta$ (Table 2). This fraction of e(a) galaxies is
similar to that in the  Hammer et al. (1995) survey, and also with the
$\sim 40$ per cent estimated by Poggianti and Wu (2000) for optically
selected merging/interacting galaxies 
(compared to 7--8 per cent in typical field galaxies). The low
$H_{\beta}/L_{8.44}$ of the e(a) galaxies might be explained if they 
are recent  post-starbursts with a small amount of  ongoing
star-formation. However,  
galaxies with active starbursts, indicated 
by $H\gamma$ emission
and/or strong oxygen lines (16V22, L21, S10), have similarly low
 $H_{\beta}/L_{8.44}$ ratios, as do the starburst Wolf-Rayet galaxies of
 Beck et al. (2000). 
This implies that, while burst-age  effects may be significant,
 dust extinction of star-forming regions
remains the  dominant effect in reducing
the average $H_{\beta}/L_{8.44}$.

Mid infra-red spectroscopy could potentially separate these effects by
providing a virtually dust-independent diagnostic  of
starburst ages
(Thornley et al. 2001), while Chandra observations of faint radio
sources could 
identify which have contributions from
obscured AGN (which produce elevated X-ray to radio flux ratios).

\section*{Appendix 1: Estimating Luminosities}
\subsection*{Optical Luminosity}  
We use the set of optical SEDs, from the  Roche et al. (2001a)
evolving models
for normal galaxies
, to  `de-k-correct' the  apparent magnitudes and  thus estimate
absolute magnitudes in the  rest-frame $B$ and $K^{\prime}$ bands. For
each   model,  we  calculate  two  observer-frame  to
restframe          corrections,         $B_{rest}-R_{obs}$         and
$K^{\prime}_{rest}-K^{\prime}_{obs}$, as  a function of  redshift. For
each galaxy  the appropriate correction  is derived from  its observed
colour, as follows.  The $B-R$  colour of each galaxy can be expressed
as a combination  of the two adjacent (i.e. one  redder and one bluer)
models at the  same redshift. If the observed colour  is $C_g$ and the
nearest  redder and  bluer models  are $C_r$  and $C_b$,  then  in the
observed   $R$   band   the   fraction   of   the   bluer   model   is
$f_{b}={10^{-0.4C_g}-10^{-0.4C_r}\over 10^{-0.4C_b}-10^{-0.4C_r}}$.

L3, for example, is intermediate  between the Im and Burst models with
$f_{b}= 0.35$.  The  restframe correction of the galaxy  $D_g$ is then
estimated by  interpolating between  the restframe corrections  of the
two models  at the same redshift,  $D_b$ and $D_r$,  as: $D_g=2.5 {\rm
log}~[f_{b}*10^{-0.4D_b}+(1-f_{b})*10^{-0.4D_r}]$.
  
This $B_{rest}-R_{obs}$  correction is then added to  the observed $R$
magnitude,    and   the    distance   modulus    (for   $\Omega_m=0.3$
$\Omega_{\Lambda}=0.7$)  is subtracted to  give an  AB-system absolute
magnitude in the rest-frame blue-band, $M_B$ (+ $5~{\rm log}~h_{50}$).
The  $R-K$ colour  of  each galaxy  can  similarly be  expressed as  a
combination of two  adjacent models, giving a ratio  of the two models
in the $K^{\prime}$-band, which  is used to interpolate,
giving  a  correction  $K^{\prime}_{rest}-K^{\prime}_{obs}$.
 This is  added to $K^{\prime}$,  and the distance
modulus subtracted, to give $M_{K}$.
\subsection*{Radio Luminosity}
The restframe
8.44 GHz luminosity (hereafter  $L_{8.44}$), in the form $\nu L_{\nu}$
ergs $\rm s^{-1}$, is estimated as

\noindent ${\rm log}~L_{8.44}=40.08-23.0+{\rm log}~(8.44\times10^9)+0.4~d_{mod}+0.4~k_{8.44}+
{\rm log}~F(8.44)$

\noindent  where  $d_{mod}$ is  the  distance  modulus in  magnitudes,
$F(8.44)$ the observed  8.44 GHz flux in $\rm  \mu Jy$, and $k_{8.44}$
the 8.44 GHz k-correction derived from the Condon (1992) radio SED,

\noindent  $k_{8.44}=-2.5~{\rm log}[(1+z)F_{\nu}(8.44[1+z]~{\rm GHz})/F_{\nu}(8.44~{\rm GHz})]$  

\noindent magnitudes.

\subsection*{Acknowledgements} This paper is based on observations
obtained at the W.M. Keck Observatory, which is operated jointly by
the University of California and the California Institute of
Technology.
We are grateful to the staff of the Keck Observatory for their expert
assistance.  Support for this work was provided by the National
Aeronautics and Space Administration through Hubble Fellowship grant
HF-1048.01-93A from the Space Telescope Science Institute, which is
operated by the Association of Universities for Research in Astronomy,
Inc., under NASA contract NAS 5-26555. 
DCK was supported by an NSF PYI grant AST-8858203 and a research
grant from UC Santa Cruz.
We thank Gabriela Mall\'{e}n-Ornelas for advice on the reduction of 
multi-object spectra. 
 
\end{document}